\documentclass[aps, prd, preprint, letterpaper, 12pt, nofootinbib, superscriptaddress, longbibliography, notitlepage, floatfix]{revtex4-2}
\usepackage{slashed}
\usepackage[normalem]{ulem}
\usepackage{amsmath,amssymb,amsfonts}
\usepackage{epsf}
\usepackage{pstricks}
\usepackage{color,xcolor}
\usepackage{multirow}
\usepackage{graphics}
\usepackage{extpfeil}
\usepackage{empheq}

\definecolor{hgreen}{rgb}{0,.7,0}
\definecolor{hred}{rgb}{.7,0,0}
\definecolor{hblue}{rgb}{0,0,.7}

\usepackage[colorlinks=true,
linkcolor=hblue,
citecolor=hgreen,
filecolor=hblue,
urlcolor=hred]{hyperref}

\allowdisplaybreaks

\begin{document}

\author{W. Altmannshofer}
\email{waltmann@ucsc.edu}
\affiliation{Department of Physics, University of California Santa Cruz,
and Santa Cruz Institute for Particle Physics,
1156 High St., Santa Cruz, CA 95064, US}

\author{S. Roy}	
\email{shibasis.cmi@gmail.com}
\affiliation{ELTE Eötvös Loránd University, Pázmány Péter 1/A, H–1117 Budapest, Hungary}

\title{\boldmath Angular Observables in the \texorpdfstring{$\Lambda_{b}\to \Lambda \nu \bar{\nu}$}{Lambda-b to Lambda nu nu-bar} Decay} 

\begin{abstract}
We study the rare decay $\Lambda_b \rightarrow \Lambda(\rightarrow p\pi^-)\nu\bar{\nu}$ of polarized $\Lambda_b$ baryons produced at the $Z$ pole. We find that the differential decay distribution is characterized by five independent angular observables, four of which require initial $\Lambda_b$ polarization. While standard observables in mesonic $b \rightarrow s \nu \bar\nu$ decays ($B \rightarrow K^{(*)} \nu \bar{\nu}$, $B_s \rightarrow \phi \nu \bar{\nu}$) probe only two combinations of short-distance Wilson coefficients, $|C_L + C_R|$ and $|C_L - C_R|$, the angular observables in $\Lambda_b \rightarrow \Lambda(\rightarrow p\pi^-)\nu\bar{\nu}$ provide access to the orthogonal combinations $|C_R|^2 - |C_L|^2$ and $\text{Im}(C_L C_R^*)$. In particular, we identify a $T$-odd CP asymmetry, $A_5$, that directly probes CP-violating phases in the short-distance effective Hamiltonian. We present Standard Model predictions for the branching ratio and all angular observables using lattice QCD form factors, and evaluate the projected new physics sensitivity at the FCC-ee. Our results demonstrate that baryonic angular observables are essential for resolving parameter degeneracies and for completely characterizing the chiral and CP structure of possible new physics in $b \rightarrow s \nu \bar{\nu}$ transitions.
\end{abstract}

\maketitle
\newpage
\tableofcontents

\section{Introduction} \label{sec:intro}

Rare flavor-changing neutral current (FCNC) transitions mediated by the quark-level process $b\to s\nu\bar\nu$ are among the theoretically cleanest flavor probes of physics beyond the Standard Model (SM). They are absent at tree level in the SM and arise only through electroweak loop diagrams, leading to highly suppressed branching ratios. At the same time, the neutrinos in the final state eliminate non-local hadronic effects that complicate the corresponding charged lepton modes, resulting in precise SM predictions. These properties make $b\to s\nu\bar\nu$ decays powerful probes of heavy new particles and non-standard interactions~\cite{Altmannshofer:2009ma, Buras:2014fpa, Browder:2021hbl, Bause:2021cna, Felkl:2021uxi, Rajeev:2021ntt,  Becirevic:2023aov}.

Experimentally, the field has recently entered a new era. Belle II has reported the first evidence for the decay $B^+\to K^+\nu\bar\nu$~\cite{Belle-II:2023esi}, marking an important milestone in the study of these elusive processes. The measured branching fraction lies about $2.7\sigma$ above the SM prediction and has motivated a variety of new physics interpretations~\cite{Athron:2023hmz, Bause:2023mfe, Allwicher:2023xba, Felkl:2023ayn, He:2023bnk, Fridell:2023ssf, Chen:2024jlj, Hou:2024vyw, Marzocca:2024hua, Rosauro-Alcaraz:2024mvx, Allwicher:2024ncl, Das:2025zrn, Prisha:2026nzr, Abada:2026dlb, Abri:2026oqv}, including light dark sectors~\cite{Wang:2023trd, Berezhnoy:2023rxx, Datta:2023iln, Altmannshofer:2023hkn, McKeen:2023uzo,  Ho:2024cwk, Gabrielli:2024wys, He:2024iju, Bolton:2024egx, Buras:2024ewl, Altmannshofer:2024kxb, Hu:2024mgf,  Calibbi:2025rpx, Berezhnoy:2025nmb, Hu:2025zua, Abumusabh:2025zsr, Kim:2025zaf, Bolton:2025lnb, Liu:2025lbw, Mescia:2026xju, Gartner:2026clx, Hong:2026qoj, Bhattacharya:2026qzt, Abumusabh:2026ykp, Ho:2026kqp, Berezhnoy:2026yii}. With its full data set, Belle II is expected to measure the branching ratios of $B\to K\nu\bar\nu$ and $B\to K^*\nu\bar\nu$ with ultimate precisions at the level of $\sim10\%$~\cite{Belle-II:2018jsg}. Looking further ahead, the clean environment and enormous $b$ hadron samples anticipated at FCC-ee promise another substantial leap in precision, with projected uncertainties of the order of $1\%$ for $B\to K\nu\bar\nu$ and $B\to K^*\nu\bar\nu$, and of the order of $10\%$ for $B_s\to\phi\nu\bar\nu$ and $\Lambda_b\to\Lambda\nu\bar\nu$~\cite{Li:2022tov, Amhis:2023mpj, AlvarezCartelle:2025mtx}.

The experimental signature of $b\to s\nu\bar\nu$ decays is intrinsically challenging because the neutrino pair escapes detection. As a result, the information that can be extracted from these decays is considerably more limited than in the corresponding charged lepton modes $b \to s \ell^+ \ell^-$, where the complete kinematics of the final state can be reconstructed. Nevertheless, valuable information remains encoded in the angular distributions of the visible decay products. Naturally, the mesonic modes will remain the flagship measurements of the $b\to s\nu\bar\nu$ program. They benefit from significantly larger event samples since $b$ quarks hadronize predominantly into $B^0$ and $B^\pm$ mesons rather than $\Lambda_b$ baryons~\cite{LHCb:2023wbo}, allowing substantially higher experimental precision. This raises the question if the baryonic decay $\Lambda_b \to \Lambda \nu\bar\nu$ can provide information beyond what is accessible in the mesonic channels.

As we emphasize in this paper, the baryonic decay $\Lambda_b \to \Lambda \nu\bar\nu$ probes aspects of the underlying short-distance physics that are not accessible in the mesonic modes in a straightforward way. The baryonic decay probes a fermion-to-fermion transition and therefore exhibits a pattern of observables that is qualitatively different from the pseudoscalar-to-pseudoscalar and pseudoscalar-to-vector transitions realized in $B\to K\nu\bar\nu$, $B\to K^*\nu\bar\nu$, and $B_s \to \phi\nu\bar\nu$. This leads to genuinely complementary sensitivity to the underlying short-distance dynamics. In particular, if deviations from the SM are observed in the mesonic channels either at Belle II or at a future FCC-ee, the measurement of $\Lambda_b\to\Lambda\nu\bar\nu$ will provide crucial additional information that helps disentangle the operator structure and chiral nature of the new interactions and could even provide information about CP violation in the $b \to s \nu\bar \nu$ decays. Indeed, we will argue that baryonic observables are important for a complete characterization of the underlying new physics. Conversely, even if future measurements of the mesonic modes remain consistent with the SM, it is possible that new physics manifests itself only in the baryonic channel. Given the theoretical cleanliness of $\Lambda_b\to\Lambda\nu\bar\nu$ and the projected experimental sensitivity at FCC-ee, there is therefore a compelling case for including this decay in the global $b\to s\nu\bar\nu$ program.

A unique feature of $\Lambda_b$ baryons produced at FCC-ee is that they inherit a sizable longitudinal polarization from the parity-violating decay $Z\to b\bar b$. This polarization gives rise to additional angular observables that are absent for unpolarized $\Lambda_b$ decays and unavailable in the corresponding mesonic decays. In~\cite{Altmannshofer:2025eor} we investigated the polarized decay $\Lambda_b\to\Lambda\nu\bar\nu$ and demonstrated that the forward-backward asymmetry of the $\Lambda$ flight direction with respect to the $\Lambda_b$ polarization axis provides qualitatively new information on the short-distance couplings. Other previous work on $\Lambda_b\to\Lambda\nu\bar\nu$ decays can be found in~\cite{Chen:2000mr, Aliev:2007rm, Sirvanli:2007yq, Das:2025zrn, Hu:2025zua}.

In the present work we extend the analysis in~\cite{Altmannshofer:2025eor} by including the subsequent parity-violating decay $\Lambda\to p\pi^-$ and study the complete cascade decay $\Lambda_b\to\Lambda(\to p\pi^-)\nu\bar\nu$. We derive the full angular distribution and show that, in addition to the branching ratio, it provides access to five independent angular observables. Four of these observables require a polarized initial $\Lambda_b$ and are therefore unique to experiments operating at the $Z$ pole, such as FCC-ee. Among the angular observables is a $T$-odd CP asymmetry that is directly sensitive to CP-violating phases in the $b\to s\nu\bar\nu$ transition. To our knowledge, the only other proposed probe of CP violation in these decays is through time-dependent CP asymmetries in flavor-tagged decays to CP eigenstates, such as $B^0\to K_S\nu\bar\nu$~\cite{Descotes-Genon:2022gcp}. The observables identified in our work therefore broaden the discovery potential of the $b\to s\nu\bar\nu$ program and enhance its ability to characterize the underlying structure of possible new physics.\footnote{For comparison, the analogous decays with charged leptons in the final state, $\Lambda_b \to \Lambda \ell^{+}\ell^{-}$, benefit from the additional angular information carried by the charged lepton pair. Such decays have been extensively investigated in the literature~\cite{Feldmann:2011xf,Detmold:2012vy,Gutsche:2013pp, Boer:2014kda, Mott:2015zma, Detmold:2016pkz, Blake:2017une, Roy:2017dum, Das:2018sms, Blake:2019guk, Blake:2022vfl, Das:2022xjg} including studies of decays into excited $\Lambda^*$ baryons~\cite{Descotes-Genon:2019dbw, Das:2020cpv, Amhis:2020phx, Meinel:2020owd, Bordone:2021bop, Meinel:2021mdj, Amhis:2022vcd, Li:2022nim, Beck:2022spd, Das:2023kch} and decays of polarized $\Lambda_b$ baryons~\cite{Das:2020qws, Beck:2025bgc}. Experimental results for the baryonic decays are available from LHCb~\cite{LHCb:2015tgy, LHCb:2017slr, LHCb:2018jna, LHCb:2019efc, LHCb:2023ptw, LHCb:2024ick, LHCb:2026dbi}. Other related work on decays of polarized $\Lambda_b$ baryons can be found here~\cite{Hrivnac:1994jx, Kramer:1996yz, Hiller:2001zj}.}

\newpage
This paper is organized as follows: In section~\ref{sec:Heff}, we introduce the effective Hamiltonian governing the $b \to s \nu \bar{\nu}$ transition in the SM and its possible extensions. In section~\ref{sec:angular}, we develop the formalism for deriving the angular distribution of the $\Lambda_b \to \Lambda (\to p \pi^-) \nu \bar{\nu}$ decay and identify all angular observables that can be measured in the decay of polarized $\Lambda_b$ baryons. In section~\ref{sec:sensitivity}, we establish the connection of the angular observables to the Wilson coefficients of the effective Hamiltonian. This section also discusses the relevant $\Lambda_b \to \Lambda$ form factors, which constitute the main source of theoretical uncertainty. We also provide SM predictions for all $\Lambda_b \to \Lambda (\to p \pi^-) \nu \bar{\nu}$ observables as well as predictions in new physics benchmark scenarios.
Section~\ref{sec:prospects} contains our main results. We determine the projected sensitivities of $\Lambda_b \to \Lambda (\to p \pi^-) \nu \bar\nu$ observables at FCC-ee both in scenarios with real and complex Wilson coefficients, monitoring the impact of the assumed experimental precision. We conclude in section~\ref{sec:conclusions}. Precise definitions of the helicity angles and the Wigner $D$ functions we use in our work are given in appendices~\ref{app:angles} and~\ref{app:wigD}. In appendix~\ref{app:q2_binning} we show how the sensitivity projections can be further improved by splitting the $q^2$ region into two bins. 

\section{Effective Hamiltonian} \label{sec:Heff}

The flavor-changing neutral-current transition $b \to s \nu \bar\nu$ is described by the effective Hamiltonian
\begin{equation} \label{eq:Heff}
\mathcal{H}_\text{eff}= - \frac{4G_{F}}{\sqrt{2}} \frac{\alpha}{4\pi} V_{ts}^*V_{tb} \Big( C_L \big(\bar s \gamma_\mu P_L b \big) + C_R \big(\bar s \gamma_\mu P_R b \big) \Big) \big(\bar \nu \gamma^\mu (1-\gamma_{5}) \nu\big) ~+~ \text{h.c.}
\end{equation}
Assuming only the three SM neutrinos and restricting ourselves to operators that arise from dimension-six operators in the Standard Model Effective Field Theory (SMEFT), the two operators shown above constitute the most general effective Hamiltonian for $b \to s \nu \bar\nu$ transitions. Scalar and tensor interactions arise only from higher-dimensional operators in SMEFT and are therefore expected to be highly suppressed. Moreover, such operators are typically strongly constrained by neutrino mass generation and will not be considered here.

For simplicity, we also restrict our discussion to SM neutrinos and do not introduce light right-handed neutrinos. Such states can modify the kinematic distributions through their non-zero masses and, being distinct particles, contribute incoherently to the decay rate. More generally, light invisible particles from a dark sector could give rise to decays such as $\Lambda_b \to \Lambda X$ or $\Lambda_b \to \Lambda XX$. While such scenarios are phenomenologically interesting, they require a dedicated analysis and lie beyond the scope of the present work. (See e.g.~\cite{Kamenik:2011vy, He:2022ljo, Wang:2023trd, Berezhnoy:2023rxx, Datta:2023iln, Altmannshofer:2023hkn, McKeen:2023uzo,  Ho:2024cwk, Gabrielli:2024wys, He:2024iju, Bolton:2024egx, Buras:2024ewl, Altmannshofer:2024kxb, Hu:2024mgf,  Calibbi:2025rpx, Berezhnoy:2025nmb, Hu:2025zua, Abumusabh:2025zsr, Kim:2025zaf, Bolton:2025lnb, Liu:2025lbw, Mescia:2026xju, Gartner:2026clx, Hong:2026qoj, Bhattacharya:2026qzt, Abumusabh:2026ykp, Ho:2026kqp, Berezhnoy:2026yii} for work on probing light invisible particles in $b \to s ~+E_\text{miss}$ decays.) 
Throughout this paper we furthermore assume that the effective interactions are flavor conserving and flavor universal in the neutrino sector. This assumption is made primarily for simplicity; none of the qualitative conclusions presented below depend on it.

The short-distance physics is encoded in the two complex Wilson coefficients $C_L$ and $C_R$. In the SM, only the left-handed operator is generated to an excellent approximation
\begin{equation}
C_L = C_L^\text{SM} + C_L^\text{NP}\,, \qquad C_R \simeq C_R^\text{NP}\,.
\end{equation}
The SM contribution is given by~\cite{Brod:2010hi, Brod:2021hsj}
\begin{equation}
C_L^\text{SM} = - \frac{1}{s_W^2} X(x_t) = - 6.305 \pm 0.027 \pm 0.073 \pm 0.009 ~,
\end{equation}
where $s_W = \sin\theta_W$ is the sine of the weak mixing angle, and the loop function $X$ depends on the ratio of top $\overline{\text{MS}}$-mass to $W$ boson mass $x_t = m_t^2/m_W^2$. The three quoted uncertainties arise from the top quark mass, unknown higher-order QCD corrections, and unknown higher-order electroweak corrections, respectively. Compared to~\cite{Brod:2021hsj} we have updated the numerical value of the Wilson coefficient with the latest result for the top quark pole mass from cross section measurements $m_t^\text{pole} = (172.1 \pm 0.6)\,\text{GeV}$~\cite{ParticleDataGroup:2024cfk}, translating it to the top $\overline{\text{MS}}$-mass using 3-loop QCD~\cite{Herren:2017osy}: $m_t = (162.5 \pm 0.6)\,\text{GeV}$ (see also~\cite{Fael:2025xmi}). 

Besides the Wilson coefficients, the effective Hamiltonian depends on the electromagnetic coupling evaluated at the electroweak scale, $\alpha^{-1} \simeq 127.95$ with negligible uncertainty, and the CKM combination $V_{ts}^* V_{tb}$. We determine the CKM factor using independent tree-level inputs from the PDG~\cite{ParticleDataGroup:2024cfk}
\begin{equation}
|V_{cb}| = (40.7 \pm 1.3)\times 10^{-3}\,,~~
|V_{ub}| = (3.89 \pm 0.16)\times 10^{-3}\,,~~
\gamma = (66.4 \pm 2.8)^\circ\,,~~
\lambda = 0.2245\,,
\end{equation}
and we neglect the tiny uncertainty in the Cabibbo angle $\lambda$. This procedure gives
\begin{equation}
    \big| V_{ts}^* V_{tb} \big|^2 =\big( 1.601 \pm 0.101 \big)\times 10^{-3} \,,
\end{equation}
where the uncertainty is entirely dominated by the present uncertainty in $|V_{cb}|$.

Since the effective Hamiltonian contains two complex Wilson coefficients, the most general short-distance physics is characterized by three independent physical parameters. Convenient choices are the absolute values of the Wilson coefficients and their relative phase:
\begin{equation}
|C_L|~,\quad |C_R| ~,\quad \phi = \text{Arg}(C_R/C_L) ~.
\end{equation} 
The overall common phase is unobservable. 

An important observation is that the standard observables in exclusive and inclusive mesonic $b\to s\nu\bar\nu$ decays, $B\to K \nu\bar\nu$, $B\to K^* \nu\bar\nu$, $B_s\to \phi \nu\bar\nu$, $B\to X_s \nu\bar\nu$, do not probe all three parameters independently. Instead, they depend only on the two combinations $|C_L+C_R|$ and $|C_L-C_R|$. This can be seen explicitly from the  expressions (based on~\cite{Altmannshofer:2009ma, Buras:2014fpa} updated with form factors from~\cite{Gubernari:2023puw})
\begin{eqnarray} \label{eq:meson1}
\frac{\text{BR}(B \to K \nu \bar\nu)}{\text{BR}(B \to K \nu \bar\nu)_\text{SM}} &=& \frac{|C_L + C_R|^2}{|C_L^\text{SM}|^2} ~,\\[8pt]   \label{eq:meson2}
\frac{\text{BR}(B \to X_s \nu \bar\nu)}{\text{BR}(B \to X_s \nu \bar\nu)_\text{SM}} &\simeq& \frac{|C_L|^2}{|C_L^\text{SM}|^2} + \frac{|C_R|^2}{|C_L^\text{SM}|^2} = \frac{1}{2} \frac{|C_L - C_R|^2}{|C_L^\text{SM}|^2} + \frac{1}{2} \frac{|C_L + C_R|^2}{|C_L^\text{SM}|^2} ~, \\[8pt]  \label{eq:meson3}
\frac{\text{BR}(B \to K^* \nu \bar\nu)}{\text{BR}(B \to K^* \nu \bar\nu)_\text{SM}} &\simeq& 0.82 \frac{|C_L - C_R|^2}{|C_L^\text{SM}|^2} + 0.18 \frac{|C_L + C_R|^2}{|C_L^\text{SM}|^2} ~,\\[8pt]  \label{eq:meson4}
\frac{F_L(B \to K^* \nu \bar\nu)}{F_L(B \to K^* \nu \bar\nu)_\text{SM}} &\simeq& \frac{|C_L - C_R|^2}{0.82 |C_L - C_R|^2 + 0.18 |C_L + C_R|^2} ~.
\end{eqnarray}
Observables accessible in $B_s \to \phi \nu \bar\nu$ have an analogous dependence on the Wilson coefficients as those in $B \to K^* \nu\bar\nu$. Consequently, even a complete measurement of branching ratios and angular observables in the mesonic channels provides only the two combinations $|C_L+C_R|$ and $|C_L-C_R|$. While the resulting overconstrained system provides powerful consistency tests of the effective Hamiltonian, it remains insensitive to the third physical parameter, the relative phase between $C_L$ and $C_R$.\footnote{An important caveat is that the relative phase can, in principle, be accessed through time-dependent CP asymmetries in flavor-tagged decays to CP eigenstates such as $B^0\to K_S\nu\bar\nu$. Although experimentally challenging, such measurements may become feasible at FCC-ee~\cite{Descotes-Genon:2022gcp}.}

One of the central points of the present work is that polarized $\Lambda_b \to \Lambda \nu \bar\nu$ decays overcome this limitation. We will show that the angular observables accessible in the baryonic decay provide sensitivity to the remaining physical parameter, thereby allowing for a complete determination of the short-distance couplings. In particular, we will identify a $T$-odd CP asymmetry that provides direct sensitivity to the CP-violating phase in the short-distance couplings.

\section{The \texorpdfstring{$\Lambda_b \to \Lambda (\to p \pi^-) \nu\bar\nu$}{Lambda-b to Lambda (to p pi-) nu nu-bar} Angular Distribution} \label{sec:angular}

In this section we discuss the angular distribution of the cascade decay $\Lambda_b \to \Lambda (\to p \pi^-) \nu\bar\nu$ of polarized $\Lambda_b$ baryons. This angular distribution forms the basis for the phenomenological analysis presented below. We begin by discussing the production of polarized $\Lambda_b$ baryons at the $Z$-pole, introduce the relevant kinematic variables, and then derive the angular distribution in the helicity formalism.

\subsection{Polarized \texorpdfstring{$\Lambda_b$}{Lambda-b} Production at the \texorpdfstring{$Z$}{Z} pole} \label{sec:Zpole}

A unique feature of the FCC-ee program is the production of polarized $\Lambda_b$ baryons at the $Z$ pole. The anticipated sample of approximately $6\times10^{12}$ $Z$ bosons will yield an enormous number of $b\bar b$ pairs~\cite{Bernardi:2022hny, FCC:2025lpp}. Owing to the parity-violating couplings of the $Z$ boson, the produced $b$ and $\bar b$ quarks are highly longitudinally polarized. During hadronization, a sizable fraction of this polarization is retained by the $\Lambda_b$ and $\bar\Lambda_b$ baryons~\cite{Mannel:1991bs, Falk:1993rf}, providing access to polarization-dependent observables that are absent in unpolarized samples.

The longitudinal polarization of $\Lambda_b$ baryons produced at the $Z$ pole has been measured at LEP by the ALEPH, OPAL, and DELPHI collaborations using semileptonic $\Lambda_b$ decays~\cite{ALEPH:1995aqx, OPAL:1998wmk, DELPHI:1999hkl}. Following~\cite{Altmannshofer:2025eor}, the naive LEP average is
\begin{equation}
\mathcal P_{\Lambda_b} = \frac{N(\lambda_{\Lambda_b}=+\frac12) - N(\lambda_{\Lambda_b}=-\frac12)}{N(\lambda_{\Lambda_b}=+\frac12) + N(\lambda_{\Lambda_b}=-\frac12)} = -0.40\pm0.14 ~,
\end{equation}
where $\lambda_{\Lambda_b}$ denotes the $\Lambda_b$ helicity, defined with respect to the $\Lambda_b$ flight direction in the laboratory frame.
Throughout this work we adopt conventions such that the polarization of $\bar\Lambda_b$ baryons is identical in size but has opposite sign to that of $\Lambda_b$ when defined with respect to each baryon's own momentum direction, $\mathcal P_{\Lambda_b} = - \mathcal P_{\bar \Lambda_b}$. This is consistent with the usual conventions used in studies of $\Lambda$ and $\bar{\Lambda}$ polarization~\cite{Chen:1994ar, ALEPH:1996oew, OPAL:1997oem}.

While initial measurements of the longitudinal polarization $\mathcal{P}_{\Lambda_b}$ at LEP were limited by statistics, the $10^5$--$10^6$ times larger sample size expected at FCC-ee should enable a very precise determination of $\mathcal{P}_{\Lambda_b}$. Consequently, the uncertainty associated with $\mathcal{P}_{\Lambda_b}$ will be negligible compared to the statistical precision achievable on the rare decay angular observables, and we treat $\mathcal{P}_{\Lambda_b}$ as fixed in our analysis.

\subsection{Definition of Angular Variables} \label{sec:angles}

\begin{figure}[tb]
    \centering
    \includegraphics[trim={1cm 19cm 1cm 2.5cm}, clip=true, width=\textwidth]{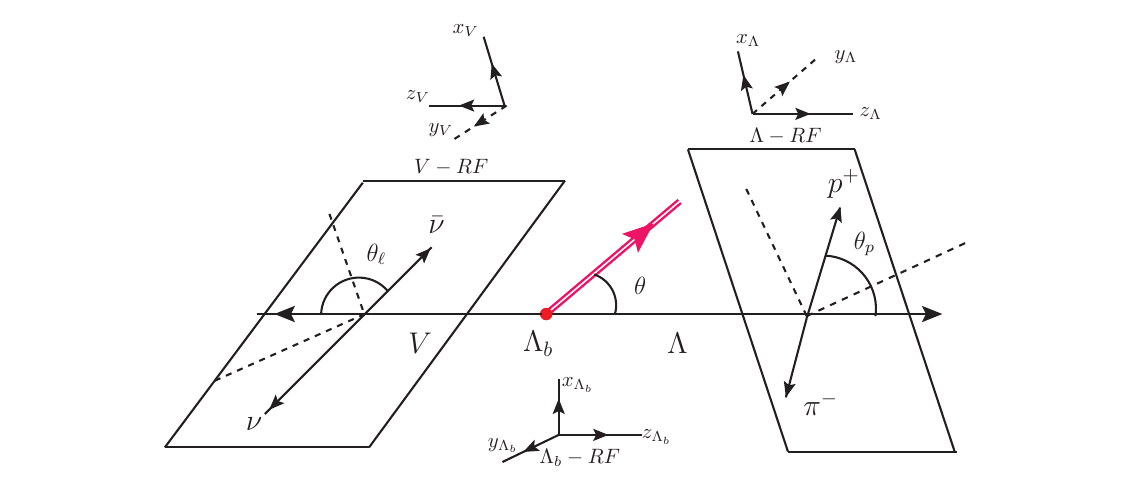}
    \caption{Illustration of the angles describing the decay $\Lambda_b\to\Lambda(\to p\pi^-)\nu\bar\nu$. See main text and appendix~\ref{app:angles} for more details.}
    \label{Fig:frames}
\end{figure}

The angular variables used throughout this work are illustrated in figure~\ref{Fig:frames}. Besides the di-neutrino invariant mass, $q^2=(p_{\Lambda_b}-p_\Lambda)^2$, the $\Lambda_b \to \Lambda (\to p \pi^-) \nu\bar\nu$ decay is described by five angles.

In the $\Lambda_b$ rest frame, the $\Lambda$ momentum defines the positive $z$-axis, while the $\Lambda_b$ polarization axis (corresponding to the $\Lambda_b$ flight direction in the lab frame) is chosen to lie in the $x$-$z$ plane. The angle $\theta$ is defined as the angle between the $\Lambda_b$ polarization axis (red in figure~\ref{Fig:frames}) and the $\Lambda$ momentum.

The proton helicity angles $(\theta_p,\phi_p)$ are defined in the $\Lambda$ rest frame with respect to the $\Lambda$ flight direction in the $\Lambda_b$ rest frame, while the antineutrino helicity angles $(\theta_{\bar \nu},\phi_{\bar \nu})$ are defined in the di-neutrino rest frame with respect to the direction of motion of the di-neutrino system in the $\Lambda_b$ rest frame. Precise definitions are given in appendix~\ref{app:angles}. Note that the neutrino angles are not observable. 

We use analogous definitions for the angles in the anti-baryon decay $\bar \Lambda_b \to \bar \Lambda (\to \bar p \pi^+) \nu \bar\nu$. In particular, we have $\bar \theta = \theta$ and $\bar \theta_p = \theta_p$. However, it will be useful to flip the sign of the azimuthal angle $\bar\phi_p = - \phi_p$ as we will see in the next section~\ref{sec:distribution}. The definition of the neutrino angles is not relevant as they are not observable.

\subsection{Angular Decay Distribution} \label{sec:distribution}

The angular distribution is conveniently derived using the helicity formalism, see e.g.~\cite{Jacob:1959at, Chung:1971ri, Richman:1984gh, Bialas:1992ny, Kramer:1996yz, Kadeer:2005aq, Gratrex:2015hna}. For a fixed helicity $\lambda_{\Lambda_b}$ of the initial $\Lambda_b$ baryon, the decay amplitude can be written as
\begin{multline}
\mathcal A(\lambda_{\Lambda_b}) \propto \sum_{X=V,A} ~ \sum_{\lambda_\Lambda=\pm \frac12} ~ \sum_{\lambda=-1,0,+1} ~
H^X_{\lambda_\Lambda,\lambda} \, D^{1/2\,*}_{\lambda_{\Lambda_b}, \lambda_\Lambda-\lambda}(0,\theta,0) \\
 \times L^X_{\lambda_{\bar\nu},\lambda_\nu} \, D^{1\,*}_{\lambda,
\lambda_{\bar\nu}-\lambda_\nu} (\phi_{\bar\nu},\theta_{\bar\nu},-\phi_{\bar\nu}) \times K_{\lambda_p} \, D^{1/2\,*}_{\lambda_\Lambda,\lambda_p} (\phi_p,\theta_p,-\phi_p)~.
\end{multline}
Here $H^X_{\lambda_\Lambda,\lambda}$, $L^X_{\lambda_{\bar\nu},\lambda_\nu}$, and $K_{\lambda_p}$ denote the helicity amplitudes of the $\Lambda_b\to\Lambda V^*$, $V^* \to \nu\bar\nu$, and $\Lambda\to p\pi^-$ transitions, and $D^{j}_{m,m^\prime}(\phi,\theta,-\phi)$ are Wigner $D$ functions that depend on the five angles introduced in section~\ref{sec:angles}. Our conventions for the Wigner $D$ functions are spelled out explicitly in appendix~\ref{app:wigD}. The index $X=V,A$ labels the vector and axial-vector components of the effective neutrino current.

The full differential decay distribution for a fixed initial $\Lambda_b$ helicity is obtained by squaring the amplitude and summing over the unobserved final-state helicities
\begin{multline}
\frac{d^6\Gamma(\lambda_{\Lambda_b})} {dq^2\,d\Omega_5} \propto \sum_{X,X^\prime=V,A} ~ \sum_{\lambda_\Lambda, \lambda_\Lambda^\prime=\pm \frac12} ~ \sum_{\lambda, \lambda^\prime =-1,0,+1} ~ \sum_{\lambda_p,\lambda_\nu, \lambda_{\bar\nu} =\pm \frac12} ~ H^X_{\lambda_\Lambda,\lambda} H^{X'\,*}_{\lambda_\Lambda',\lambda'} ~ L^X_{\lambda_{\bar\nu},\lambda_\nu} L^{X'\,*}_{\lambda_{\bar\nu},\lambda_\nu} ~ \big| K_{\lambda_p} \big|^2 \\
\times D^{1/2\,*}_{\lambda_{\Lambda_b}, \lambda_\Lambda-\lambda}(0,\theta,0) D^{1/2}_{\lambda_{\Lambda_b}, \lambda_\Lambda'-\lambda'}(0,\theta,0) \\
\times D^{1\,*}_{\lambda, \lambda_{\bar\nu}-\lambda_\nu} (\phi_{\bar\nu},\theta_{\bar\nu},-\phi_{\bar\nu}) D^{1}_{\lambda', \lambda_{\bar\nu}-\lambda_\nu}(\phi_{\bar\nu},\theta_{\bar\nu},-\phi_{\bar\nu}) \\
\times D^{1/2\,*}_{\lambda_\Lambda,\lambda_p}(\phi_p,\theta_p,-\phi_p) D^{1/2}_{\lambda_\Lambda',\lambda_p} (\phi_p,\theta_p,-\phi_p)~,
\end{multline}
where $d\Omega_5 = d\cos\theta\,d\cos\theta_p \,d\phi_p \, d\cos\theta_{\bar\nu} \,d\phi_{\bar\nu}$. 
Concerning the helicity amplitudes of the $\Lambda_b \to \Lambda V^*$ decay, it is convenient to express them in terms of transversity amplitudes. The non-vanishing ones are~\cite{Blake:2017une}
\begin{eqnarray}
H^V_{+\frac12,0} = - H^A_{+\frac12,0} = \frac{1}{2} \big( A_{\parallel\,0}^L + A_{\perp\,0}^L \big) ~,&& \quad H^V_{-\frac12,0} = - H^A_{-\frac12,0} = \frac{1}{2} \big( A_{\parallel\,0}^L - A_{\perp\,0}^L \big) ~, \\
H^V_{+\frac12,+1} = - H^A_{+\frac12,+1} = \frac{1}{2} \big( A_{\parallel\,1}^L + A_{\perp\,1}^L \big) ~,&& \quad H^V_{-\frac12,-1} = - H^A_{-\frac12,-1} = \frac{1}{2} \big( A_{\parallel\,1}^L - A_{\perp\,1}^L \big) ~.
\end{eqnarray}

The lepton helicity amplitudes have a very simple form. Neglecting the neutrino masses (which is an excellent approximation) one has~\cite{Gratrex:2015hna}
\begin{equation}
L^{V}_{-\frac12,+\frac12} = L^{V}_{+\frac12,-\frac12} = -\sqrt{2 q^2} ~,\quad L^{A}_{-\frac12,+\frac12} = -L^{A}_{+\frac12,-\frac12} = \sqrt{2 q^2} ~,
\end{equation}
and all the other leptonic helicity amplitudes vanish.

The parity-violating decay $\Lambda\to p\pi^-$ serves as an analyzer of the $\Lambda$ polarization. The corresponding helicity amplitudes are conveniently parameterized in terms of the decay asymmetry parameter $\alpha_\Lambda$ in the following way
\begin{equation}
|K_{+\frac12}|^2 \propto \frac12(1+\alpha_\Lambda) ~,\quad |K_{-\frac12}|^2 \propto \frac12(1-\alpha_\Lambda) ~.
\end{equation}
Assuming negligible CP violation in the weak decays $\Lambda\to p\pi^-$ and $\bar\Lambda\to\bar p\pi^+$, the decay asymmetries satisfy $\alpha_{\bar\Lambda}=-\alpha_\Lambda$, where the minus sign follows from the conventional definition of the asymmetry parameter. Experimentally, the asymmetries have been measured with excellent precision. The current PDG averages are~\cite{ParticleDataGroup:2024cfk}
\begin{equation}
\alpha_\Lambda=0.749\pm0.008 ~,\quad \alpha_{\bar\Lambda}=-0.758\pm0.005 ~.
\end{equation}
A naive combination under the assumption of CP conservation yields $\alpha_\Lambda=-\alpha_{\bar\Lambda}=0.755\pm0.004$, which we use throughout our numerical analysis.

As discussed previously in section~\ref{sec:Zpole}, the experimentally produced $\Lambda_b$ sample at the $Z$ pole is not fully polarized. The longitudinal polarization $\mathcal P_{\Lambda_b}$ corresponds to fractions $f_{\pm}=(1\pm \mathcal P_{\Lambda_b})/2$ of $\Lambda_b$ baryons with helicity
$\lambda_{\Lambda_b}=\pm1/2$, respectively. 
The angular distribution for the polarized ensemble is therefore obtained from
\begin{equation}
\frac{d^6\Gamma} {dq^2\,d\Omega_5} = \frac12(1+ \mathcal P_{\Lambda_b})
\frac{d^6\Gamma(\lambda_{\Lambda_b} = +\frac12)}{dq^2\,d\Omega_5} + \frac12(1- \mathcal P_{\Lambda_b})
\frac{d^6\Gamma(\lambda_{\Lambda_b} = -\frac12)}{dq^2\,d\Omega_5} ~.
\end{equation}
As the neutrino helicity angles $(\theta_{\bar\nu},\phi_{\bar\nu})$ cannot be reconstructed experimentally, we integrate over these variables. The experimentally accessible angular distribution then depends only on the hadronic angles $(\theta,\theta_p,\phi_p)$. We find the following form of the decay distributions for the $\Lambda_b$ and $\bar \Lambda_b$ decays
\begin{multline}
\frac{d^4 \Gamma}{dq^2 d\Omega_3} = \frac{1}{8\pi} \frac{d\Gamma}{dq^2} \bigg( 1 + \alpha_\Lambda S_1 \cos\theta_p + \mathcal P_{\Lambda_b} S_2 \cos\theta + \alpha_\Lambda \mathcal P_{\Lambda_b} S_3 \cos\theta_p \cos\theta \\
+ \alpha_\Lambda \mathcal P_{\Lambda_b} \Big( S_4 \cos\phi_p + A_5 \sin\phi_p \Big) \sin\theta_p \sin\theta \bigg) ~,
\end{multline}
\begin{multline}
\frac{d^4 \bar\Gamma}{dq^2 d\Omega_3} = \frac{1}{8\pi} \frac{d\bar\Gamma}{dq^2} \bigg( 1 - \alpha_{\bar\Lambda} S_1 \cos\bar\theta_p - \mathcal P_{\bar\Lambda_b} S_2 \cos\bar\theta + \alpha_{\bar\Lambda} \mathcal P_{\bar\Lambda_b} S_3 \cos\bar\theta_p \cos\bar\theta \\
+ \alpha_{\bar\Lambda} \mathcal P_{\bar\Lambda_b} \Big( S_4 \cos\bar\phi_p - A_5 \sin\bar\phi_p \Big) \sin\bar\theta_p \sin\bar\theta \bigg) ~,
\end{multline}
As discussed in section~\ref{sec:angles} we use the following angular conventions for the $\bar \Lambda_b$ decay compared to the $\Lambda_b$ decay: $\bar \theta = \theta$, $\bar \theta_p = \theta_p$, and $\bar \phi_p = - \phi_p$. Due to the sign change in the azimuthal angle, we can simply add the $\Lambda_b$ and $\bar\Lambda_b$ decay distributions and obtain
\begin{multline}
\frac{d^4 \big( \Gamma + \bar \Gamma \big)}{dq^2 d\Omega_3} = \frac{1}{8\pi} \frac{d\big( \Gamma + \bar \Gamma \big)}{dq^2} \bigg( 1 + \alpha_\Lambda S_1 \cos\theta_p + \mathcal P_{\Lambda_b} S_2 \cos\theta + \alpha_\Lambda \mathcal P_{\Lambda_b} S_3 \cos\theta_p \cos\theta \\
+ \alpha_\Lambda \mathcal P_{\Lambda_b} \Big( S_4 \cos\phi_p + A_5 \sin\phi_p \Big) \sin\theta_p \sin\theta \bigg) ~,
\end{multline}
where $d\Omega_3=d\cos\theta\,d\cos\theta_p\,d\phi_p$. The angular distribution is thus completely characterized by five independent observables, which we denote by $S_1$, $S_2$, $S_3$, $S_4$, and $A_5$. The chosen notation emphasizes that the four quantities $S_i$ are CP-even observables, while $A_5$ is a $T$-odd CP asymmetry that is proportional to the sine of a weak phase. Owing to the absence of strong phases in $b\to s\nu\bar\nu$ decays, there are no direct CP asymmetries, implying that the differential rate difference $d\big(\Gamma-\bar\Gamma\big)$ vanishes both in the SM and in new physics scenarios.

The individual angular observables can be extracted from the angular distribution in the following way
\begin{eqnarray}
\alpha_\Lambda S_1 &=& 2 \left( \int_0^1 - \int_{-1}^0 \right)d\cos\theta_p ~ \frac{d^2(\Gamma + \bar \Gamma)}{dq^2 \,d\cos\theta_p} \bigg/  \frac{d(\Gamma + \bar \Gamma)}{dq^2} ~, \\ 
\mathcal P_{\Lambda_b} S_2 &=& 2 \left( \int_0^1 - \int_{-1}^0 \right)d\cos\theta ~ \frac{d^2(\Gamma + \bar \Gamma)}{dq^2 \,d\cos\theta} \bigg/  \frac{d(\Gamma + \bar \Gamma)}{dq^2} ~, \\ 
\alpha_\Lambda \mathcal P_{\Lambda_b} S_3 &=& 4 \left( \int_0^1 - \int_{-1}^0 \right)d\cos\theta_p \left( \int_0^1 - \int_{-1}^0 \right)d\cos\theta~ \frac{d^3(\Gamma + \bar \Gamma)}{dq^2 \,d\cos\theta \,d\cos\theta_p } \bigg/  \frac{d(\Gamma + \bar \Gamma)}{dq^2}~, \\ 
\alpha_\Lambda \mathcal P_{\Lambda_b} S_4 &=& \frac{8}{\pi} \left( \int_0^{\pi/2} - \int_{\pi/2}^{3\pi/2} + \int_{3\pi/2}^{2\pi} \right)d\phi_p ~ \frac{d^2(\Gamma + \bar \Gamma)}{dq^2 \,d\phi_p} \bigg/  \frac{d(\Gamma + \bar \Gamma)}{dq^2}~, \\ 
\alpha_\Lambda \mathcal P_{\Lambda_b} A_5 &=& \frac{8}{\pi} \left( \int_0^{\pi} - \int_{\pi}^{2\pi} \right)d\phi_p ~ \frac{d^2(\Gamma + \bar \Gamma)}{dq^2 \,d\phi_p} \bigg/  \frac{d(\Gamma + \bar \Gamma)}{dq^2} ~.
\end{eqnarray}

The observable $S_1$ corresponds to the hadronic forward-backward asymmetry $A_\text{FB}^\Lambda$~\cite{Boer:2014kda, Amhis:2023mpj}, while $S_2$ corresponds to the spin forward-backward asymmetry $A_\text{FB}^\uparrow$, first introduced in~\cite{Altmannshofer:2025eor}
\begin{equation} \label{eq:AFB}
 A_\text{FB}^\Lambda = \frac{\alpha_\Lambda}{2} S_1 ~,\quad A_\text{FB}^\uparrow = \frac{\mathcal P_{\Lambda_b}}{2} S_2 ~. 
\end{equation}
The observable $S_1$ is measurable even when the initial $\Lambda_{b}$ is unpolarized and depends only on the decay asymmetry parameter of the subsequent parity-violating decay of the $\Lambda$. Its counterpart in the decay to charged leptons $\Lambda_b \to \Lambda \ell^+ \ell^-$ is known as an important probe of new physics~\cite{Roy:2017dum}. All other angular observables, $S_2$, $S_3$, $S_4$, and $A_5$ do depend on the $\Lambda_b$ polarization $\mathcal P_{\Lambda_b}$. 

In terms of the transversity amplitudes we find
\begin{eqnarray}
\frac{d\big( \Gamma + \bar \Gamma \big)}{dq^2} &\propto& | A_{\perp\,0}^L|^2 + | A_{\parallel\,0}^L|^2 + | A_{\perp\,1}^L|^2 + | A_{\parallel\,1}^L|^2 ~,\\[8pt]
S_1 &=& \frac{2\,\text{Re}\Big( A_{\perp\,0}^L A_{\parallel\,0}^{L\,*} + A_{\perp\,1}^L A_{\parallel\,1}^{L\,*} \Big)}{| A_{\perp\,0}^L|^2 + | A_{\parallel\,0}^L|^2 + | A_{\perp\,1}^L|^2 + | A_{\parallel\,1}^L|^2} ~, \\[8pt]
S_2 &=& \frac{2\,\text{Re}\Big( A_{\perp\,0}^L A_{\parallel\,0}^{L\,*} - A_{\perp\,1}^L A_{\parallel\,1}^{L\,*} \Big)}{| A_{\perp\,0}^L|^2 + | A_{\parallel\,0}^L|^2 + | A_{\perp\,1}^L|^2 + | A_{\parallel\,1}^L|^2} ~, \\[8pt]
S_3 &=& \frac{| A_{\perp\,0}^L|^2 + | A_{\parallel\,0}^L|^2 - | A_{\perp\,1}^L|^2 - | A_{\parallel\,1}^L|^2}{| A_{\perp\,0}^L|^2 + | A_{\parallel\,0}^L|^2 + | A_{\perp\,1}^L|^2 + | A_{\parallel\,1}^L|^2} ~, \\[8pt]
S_4 &=& \frac{| A_{\perp\,0}^L|^2 - | A_{\parallel\,0}^L|^2}{| A_{\perp\,0}^L|^2 + | A_{\parallel\,0}^L|^2 + | A_{\perp\,1}^L|^2 + | A_{\parallel\,1}^L|^2} ~, \\[8pt]
A_5 &=& \frac{2 \, \text{Im} \Big( A_{\perp\,0}^L A_{\parallel\,0}^{L\,*} \Big)}{| A_{\perp\,0}^L|^2 + | A_{\parallel\,0}^L|^2 + | A_{\perp\,1}^L|^2 + | A_{\parallel\,1}^L|^2} ~.
\end{eqnarray}
Our results are consistent with the expressions given in~\cite{Blake:2017une}.

\section{Theory Predictions for \texorpdfstring{$\Lambda_b \to \Lambda (\to p \pi^-) \nu\bar\nu$}{Lambda-b to Lambda (to p pi-) nu nu-bar} Decays} \label{sec:sensitivity}

In this section we connect the angular observables derived above to the underlying short-distance dynamics. We express the transversity amplitudes in terms of the $\Lambda_b\to\Lambda$ form factors and the Wilson coefficients of the effective Hamiltonian. This allows us to identify which combinations of short-distance couplings are probed by the different observables and, in particular, which observables provide access to the relative phase between $C_L$ and $C_R$. We then derive the SM predictions and illustrate the sensitivity of the observables to new physics.

\subsection{Hadronic Matrix Elements and Form Factors} \label{sec:formfactors}

The hadronic matrix elements entering the $\Lambda_b \to \Lambda \nu\bar\nu$ decay are parameterized in terms of vector and axial-vector form factors as~\cite{Feldmann:2011xf, Detmold:2016pkz, Blake:2022vfl}
\begin{multline}
\langle \Lambda | \bar s \gamma^\mu b |\Lambda_b \rangle  = \bar u_\Lambda \Bigg[ f_t^V(q^2) (m_{\Lambda_b} - m_\Lambda) \frac{q^\mu}{q^2} + f_\perp^V(q^2) \left( \gamma^\mu - \frac{2 (m_\Lambda p_{\Lambda_b}^\mu + m_{\Lambda_b} p_{\Lambda}^\mu)}{(m_{\Lambda_b} + m_\Lambda)^2 - q^2} \right)  \\
+ f_0^V(q^2) \frac{m_{\Lambda_b} + m_\Lambda}{(m_{\Lambda_b} + m_\Lambda)^2 - q^2} \left( p_{\Lambda_b}^\mu + p_\Lambda^\mu - (m_{\Lambda_b}^2 - m_\Lambda^2) \frac{q^\mu}{q^2} \right) \Bigg] u_{\Lambda_b}~,
\end{multline}
\begin{multline}
\langle \Lambda | \bar s \gamma^\mu \gamma_5 b |\Lambda_b \rangle  = - \bar u_\Lambda \gamma_5 \Bigg[ f_t^A(q^2) (m_{\Lambda_b} + m_\Lambda) \frac{q^\mu}{q^2} + f_\perp^A(q^2) \left( \gamma^\mu + \frac{2 (m_\Lambda p_{\Lambda_b}^\mu - m_{\Lambda_b} p_\Lambda^\mu)}{(m_{\Lambda_b} - m_\Lambda)^2 - q^2} \right)  \\
+ f_0^A(q^2) \frac{m_{\Lambda_b} - m_\Lambda}{(m_{\Lambda_b} - m_\Lambda)^2 - q^2} \left( p_{\Lambda_b}^\mu + p_\Lambda^\mu - (m_{\Lambda_b}^2 - m_\Lambda^2) \frac{q^\mu}{q^2} \right) \Bigg] u_{\Lambda_b}~,
\end{multline}
where $p_{\Lambda_b}$ and $p_\Lambda$ denote the 4-momenta of the initial-state $\Lambda_b$ and the final-state $\Lambda$, respectively, and $m_{\Lambda_b}$ and $m_\Lambda$ are the corresponding baryon masses. The form factors $f_{t,0,\perp}^{V,A}$ are functions of the momentum transfer squared, $q^2=(p_{\Lambda_b}-p_\Lambda)^2$, which is equal to the invariant mass squared of the neutrino pair. Throughout this work, we use the lattice QCD determination of the $\Lambda_b\to\Lambda$ form factors from~\cite{Detmold:2016pkz}.

Using these form factors, the transversity amplitudes introduced in section~\ref{sec:distribution} can be written in the following way
\begin{eqnarray}
 A_{\parallel\,0}^L &=& - 2 \sqrt{2} \big( C_L + C_R \big) f_0^V(q^2) \frac{m_{\Lambda_b}^2}{\sqrt{q^2}} \left( 1 + \frac{m_\Lambda}{m_{\Lambda_b}}\right) \sqrt{ \left(1 - \frac{m_\Lambda}{m_{\Lambda_b}} \right)^2 - \frac{q^2}{m_{\Lambda_b}^2}}~, \\
 A_{\perp\,0}^L &=& 2 \sqrt{2} \big( C_L - C_R \big) f_0^A(q^2) \frac{m_{\Lambda_b}^2}{\sqrt{q^2}} \left( 1 - \frac{m_\Lambda}{m_{\Lambda_b}}\right) \sqrt{ \left(1 + \frac{m_\Lambda}{m_{\Lambda_b}} \right)^2 - \frac{q^2}{m_{\Lambda_b}^2}}~, \\
 A_{\parallel\,1}^L &=& -4\big( C_L + C_R \big) f_\perp^V(q^2) m_{\Lambda_b} \sqrt{ \left(1 - \frac{m_\Lambda}{m_{\Lambda_b}} \right)^2 - \frac{q^2}{m_{\Lambda_b}^2}} ~, \\
 A_{\perp\,1}^L &=& 4\big( C_L - C_R \big) f_\perp^A(q^2) m_{\Lambda_b} \sqrt{ \left(1 + \frac{m_\Lambda}{m_{\Lambda_b}} \right)^2 - \frac{q^2}{m_{\Lambda_b}^2}} ~.
\end{eqnarray}
Note that here we do not follow the conventions in~\cite{Boer:2014kda} but the ones in~\cite{Blake:2017une}, such that the amplitudes $A^L_{\parallel\,0}$ and $A^L_{\parallel\,1}$ contain the matrix elements of quark vector currents, while $A^L_{\perp\,0}$ and $A^L_{\perp\,1}$ contain the matrix elements of axial-vector currents.

It turns out to be useful to introduce the following shorthand notation for combinations of form factors that will appear later on
\begin{eqnarray}
 \mathcal F_0^V &=& \big[f_0^V(q^2)\big]^2 \left( 1 + \frac{m_\Lambda}{m_{\Lambda_b}} \right)^2 \left(  \left( 1 - \frac{m_\Lambda}{m_{\Lambda_b}} \right)^2 - \frac{q^2}{m_{\Lambda_b}^2}\right) ~, \\
 \mathcal F_\perp^V &=& \big[ f_\perp^V(q^2)\big]^2  \frac{2q^2}{m_{\Lambda_b}^2} \left(  \left( 1 - \frac{m_\Lambda}{m_{\Lambda_b}} \right)^2 - \frac{q^2}{m_{\Lambda_b}^2}\right) ~, \\
 \mathcal F_0^A &=& \big[f_0^A(q^2)\big]^2 \left( 1 - \frac{m_\Lambda}{m_{\Lambda_b}} \right)^2 \left(  \left( 1 + \frac{m_\Lambda}{m_{\Lambda_b}} \right)^2 - \frac{q^2}{m_{\Lambda_b}^2}\right) ~, \\
 \mathcal F_\perp^A &=& \big[ f_\perp^A(q^2)\big]^2  \frac{2q^2}{m_{\Lambda_b}^2} \left(  \left( 1 + \frac{m_\Lambda}{m_{\Lambda_b}} \right)^2 - \frac{q^2}{m_{\Lambda_b}^2}\right) ~, \\
 \mathcal F_0^{VA} &=& f_0^V(q^2) f_0^A(q^2)  \left( 1 - \frac{m_\Lambda^2}{m_{\Lambda_b}^2} \right) \sqrt{\lambda\left( 1, \frac{m_\Lambda^2}{m_{\Lambda_b}^2}, \frac{q^2}{m_{\Lambda_b}^2}\right)}~, \\
 \mathcal F_\perp^{VA} &=& f_\perp^V(q^2) f_\perp^A(q^2) \frac{2q^2}{m_{\Lambda_b}^2} \sqrt{\lambda\left( 1, \frac{m_\Lambda^2}{m_{\Lambda_b}^2}, \frac{q^2}{m_{\Lambda_b}^2}\right)} ~,
\end{eqnarray}
where $\lambda(x,y,z) = x^2 + y^2 + z^2 - 2 (xy+xz+yz)$. Note that we use a slightly different notation compared to~\cite{Altmannshofer:2025eor}.

\subsection{Dependence of the Observables on the Wilson Coefficients} \label{sec:Wilson}

Using the notation for the form factor combinations introduced above, we find the following expressions for the $\Lambda_b \to \Lambda \nu\bar\nu$ branching ratio and the five angular observables 
\begin{eqnarray} \label{eq:baryon1}
\frac{d \text{BR}}{dq^2} &=& \tau_{\Lambda_b} \frac{\alpha^2 G_F^2}{256 \pi^5} m_{\Lambda_b}^3 |V_{ts}^* V_{tb}|^2 \sqrt{\lambda\left( 1, \frac{m_\Lambda^2}{m_{\Lambda_b}^2}, \frac{q^2}{m_{\Lambda_b}^2}\right)} \nonumber \\ 
&& \qquad\qquad \times \bigg[ \big| C_R + C_L \big|^2 \Big( \mathcal F_0^V + \mathcal F_\perp^V \Big) + \big| C_R - C_L \big|^2 \Big( \mathcal F_0^A + \mathcal F_\perp^A \Big) \bigg] ~, \\[8pt]  \label{eq:baryon2}
S_1 &=& \frac{2 \big(\big| C_R \big|^2 - \big| C_L \big|^2 \big) \Big( \mathcal F_0^{VA} + \mathcal F_\perp^{VA} \Big)}{\big| C_R + C_L \big|^2 \Big( \mathcal F_0^V + \mathcal F_\perp^V \Big) + \big| C_R - C_L \big|^2 \Big( \mathcal F_0^A + \mathcal F_\perp^A \Big)} ~, \\[8pt]  \label{eq:baryon3}
S_2 &=& \frac{2 \big(\big| C_R \big|^2 - \big| C_L \big|^2 \big) \Big( \mathcal F_0^{VA} - \mathcal F_\perp^{VA} \Big)}{\big| C_R + C_L \big|^2 \Big( \mathcal F_0^V + \mathcal F_\perp^V \Big) + \big| C_R - C_L \big|^2 \Big( \mathcal F_0^A + \mathcal F_\perp^A \Big)} ~, \\[8pt]  \label{eq:baryon4}
S_3 &=& \frac{\big| C_R + C_L \big|^2 \Big( \mathcal F_0^V - \mathcal F_\perp^V \Big) + \big| C_R - C_L \big|^2 \Big( \mathcal F_0^A - \mathcal F_\perp^A \Big)}{\big| C_R + C_L \big|^2 \Big( \mathcal F_0^V + \mathcal F_\perp^V \Big) + \big| C_R - C_L \big|^2 \Big( \mathcal F_0^A + \mathcal F_\perp^A \Big)} ~, \\[8pt]  \label{eq:baryon5}
S_4 &=& \frac{\big| C_R - C_L \big|^2 \mathcal F_0^A -\big| C_R + C_L \big|^2 \mathcal F_0^V}{\big| C_R + C_L \big|^2 \Big( \mathcal F_0^V + \mathcal F_\perp^V \Big) + \big| C_R - C_L \big|^2 \Big( \mathcal F_0^A + \mathcal F_\perp^A \Big)} ~, \\[8pt]  \label{eq:baryon6}
A_5 &=& \frac{-4 \text{Im} \big(C_L C_R^* \big) \mathcal F_0^{VA}}{\big| C_R + C_L \big|^2 \Big( \mathcal F_0^V + \mathcal F_\perp^V \Big) + \big| C_R - C_L \big|^2 \Big( \mathcal F_0^A + \mathcal F_\perp^A \Big)} ~.
\end{eqnarray}
All the above observables are functions of the di-neutrino invariant mass squared $q^2$.
We also introduce $q^2$ integrated versions of the branching ratio and the angular observables
\begin{equation}
\langle \text{BR} \rangle = \int dq^2~ \frac{d\text{BR}}{dq^2} ~, ~~~ \langle S_i \rangle = \frac{1}{\langle \text{BR} \rangle} \int dq^2~ \frac{d\text{BR}}{dq^2} S_i ~,~~~ \langle A_5 \rangle = \frac{1}{\langle \text{BR} \rangle} \int dq^2~ \frac{d\text{BR}}{dq^2} A_5 ~.
\end{equation}

We emphasize that the $\Lambda_b\to\Lambda\nu\bar\nu$ decay provides genuinely complementary information to the mesonic $b\to s\nu\bar\nu$ modes. The branching ratio as well as the angular observables $S_3$ and $S_4$ depend only on the two combinations $|C_L+C_R|$ and $|C_L-C_R|$, just like the observables in $B\to K\nu\bar\nu$, $B\to K^*\nu\bar\nu$, $B_s\to\phi\nu\bar\nu$, and $B\to X_s\nu\bar\nu$, cf. equations~\eqref{eq:meson1}-\eqref{eq:meson4}.\footnote{This behavior can be exploited to formulate sum rules between mesonic observables and the baryonic branching ratio, see~\cite{Kitahara:2026doj}.} By contrast, the observables $S_1$, $S_2$, and $A_5$ are sensitive to the orthogonal combinations $|C_R|^2-|C_L|^2$ and $\text{Im}(C_L C_R^*)$, which cannot be accessed through standard mesonic observables.

We note that the four combinations of Wilson coefficients appearing in the mesonic and baryonic observables are not independent, but satisfy the identity
\begin{equation}
\Big(|C_R|^2-|C_L|^2\Big)^2 + \Big(2\,\mathrm{Im}(C_LC_R^*)\Big)^2 = |C_R+C_L|^2\,|C_R-C_L|^2~.
\end{equation}
Precision measurements of mesonic observables determine the quantity on the right-hand side, thereby fixing the radius of a circle in the plane spanned by $|C_R|^2-|C_L|^2$ and $2\,\mathrm{Im}(C_LC_R^*)$. Uncertainties broaden this circle into a ring-shaped allowed region for the short-distance couplings, leaving a continuous degeneracy. Measurements of the baryonic observables $S_1$, $S_2$, and $A_5$ determine the position on this ring, removing the degeneracy and allowing for a complete determination of the underlying Wilson coefficients, up to an unobservable overall phase.

\subsection{Standard Model Predictions} \label{sec:SM}

Having established the dependence of the observables on the Wilson coefficients, we next discuss their predictions within the SM. In the SM, only the left-handed operator contributes, with $C_L=C_L^\text{SM}$ and $C_R \simeq 0$. The differential branching ratio can therefore be written as
\begin{equation}
\frac{d \text{BR}_\text{SM}}{dq^2} = \tau_{\Lambda_b} \frac{\alpha^2 G_F^2}{256 \pi^5} m_{\Lambda_b}^3 |V_{ts}^* V_{tb}|^2 \sqrt{\lambda\left( 1, \frac{m_\Lambda^2}{m_{\Lambda_b}^2}, \frac{q^2}{m_{\Lambda_b}^2}\right)} \big| C_L^\text{SM} \big|^2 \Big( \mathcal F_0^V + \mathcal F_\perp^V +\mathcal F_0^A + \mathcal F_\perp^A \Big) ~.
\end{equation}
In the CP averaged angular observables $S_1$, $S_2$, $S_3$, and $S_4$, the Wilson coefficient cancels in the SM, leaving only combinations of form factors 
\begin{eqnarray}
S_1^\text{SM} = \frac{-2 \big( \mathcal F_0^{VA} + \mathcal F_\perp^{VA} \big)}{\mathcal F_0^V + \mathcal F_\perp^V + \mathcal F_0^A + \mathcal F_\perp^A } ~, && \qquad S_2^\text{SM} = \frac{-2 \big( \mathcal F_0^{VA} - \mathcal F_\perp^{VA} \big)}{\mathcal F_0^V + \mathcal F_\perp^V + \mathcal F_0^A + \mathcal F_\perp^A }~, \\[8pt] 
S_3^\text{SM} = \frac{\mathcal F_0^V - \mathcal F_\perp^V + \mathcal F_0^A - \mathcal F_\perp^A}{\mathcal F_0^V + \mathcal F_\perp^V + \mathcal F_0^A + \mathcal F_\perp^A } ~, && \qquad S_4^\text{SM} = \frac{\mathcal F_0^A - \mathcal F_0^V}{\mathcal F_0^V + \mathcal F_\perp^V + \mathcal F_0^A + \mathcal F_\perp^A }~. 
\end{eqnarray}
Finally, due to the absence of right-handed currents and a CP-violating phase, the CP asymmetry $A_5$ vanishes in the SM
\begin{equation}
A_5^\text{SM} =0 ~.
\end{equation}

Integrating over the entire available $q^2$ range $0 < q^2 < (m_{\Lambda_b} - m_\Lambda)^2 \simeq 20.3\,\text{GeV}^2$, we obtain the following SM prediction for the $\Lambda_b \to \Lambda \nu\bar\nu$ branching ratio
\begin{equation} \label{eq:SM_prediction_BR}
\langle\text{BR}\rangle(\Lambda_b \to \Lambda \nu\bar\nu)_\text{SM} = (7.59  \pm 1.03) \times 10^{-6}  ~,
\end{equation}
which updates the result from~\cite{Altmannshofer:2025eor} using the current values for the top mass, the CKM input (see section~\ref{sec:Heff}), and the $\Lambda_b$ lifetime, $\tau_{\Lambda_b} = (1.465 \pm 0.009)\,\text{ps}$~\cite{ParticleDataGroup:2024cfk}. The dominant uncertainty is due to the form factors, followed by CKM input. The other sources of uncertainty are much smaller
\begin{multline}
 \delta\langle\text{BR}\rangle(\Lambda_b \to \Lambda \nu\bar\nu)_\text{SM} \times 10^6 = \pm 0.89 \Big|_\text{ff} \pm 0.48 \Big|_\text{CKM} \pm 0.18 \Big|_\text{QCD} \\ \pm 0.06 \Big|_{m_t} \pm 0.05 \Big|_{\tau_{\Lambda_b}} \pm 0.02 \Big|_\text{EW} = \pm 1.03~.
\end{multline}

The only sources of uncertainty in the angular observables are the form factors. We find the SM predictions
\begin{eqnarray}
 \langle S_1^\text{SM} \rangle = - 0.926 \pm 0.018 ~,&&\quad   \langle S_2^\text{SM} \rangle = - 0.053 \pm 0.068 ~, \\
 \langle S_3^\text{SM} \rangle = 0.025 \pm 0.069 ~,&&\quad   \langle S_4^\text{SM} \rangle = 0.029 \pm 0.047 ~.
\end{eqnarray}
The integrated values of $S_2$, $S_3$, and $S_4$ are close to zero due to cancellations over the full $q^2$ range. As discussed below, these cancellations originate from zero crossings in the corresponding $q^2$ distributions. By contrast, $S_1$ does not feature a zero crossing and integrates to a sizable result. Moreover, the uncertainty of $\langle S_1^\text{SM} \rangle$ turns out to be particularly small, suggesting it is an observable highly sensitive to new physics.

The uncertainties of the integrated branching ratio $\langle\text{BR}\rangle(\Lambda_b \to \Lambda \nu\bar\nu)_\text{SM}$ and the integrated angular observables $\langle S_1^\text{SM} \rangle$,  $\langle S_2^\text{SM} \rangle$,  $\langle S_3^\text{SM} \rangle$,  and $\langle S_4^\text{SM} \rangle$ are correlated. We find the correlation matrix
\begin{equation}
 \rho = \begin{pmatrix} 
 1 & -0.168 & - 0.168 & 0.162 & 0.034 \\
 -0.168 & 1 & 0.262 & -0.345 & 0.246 \\
 -0.168 & 0.262 & 1 & -0.987 & 0.139 \\
 0.162 & -0.345 & -0.987 & 1 & -0.172 \\
 0.034 & 0.246 & 0.139 & -0.172 & 1 \end{pmatrix} ~.
\end{equation}
We observe a very strong anti-correlation between the uncertainties of $\langle S_2^\text{SM} \rangle$ and $\langle S_3^\text{SM} \rangle$. The remaining correlations are modest.

\begin{figure}[tb]
\centering
\includegraphics[width=0.46\textwidth]{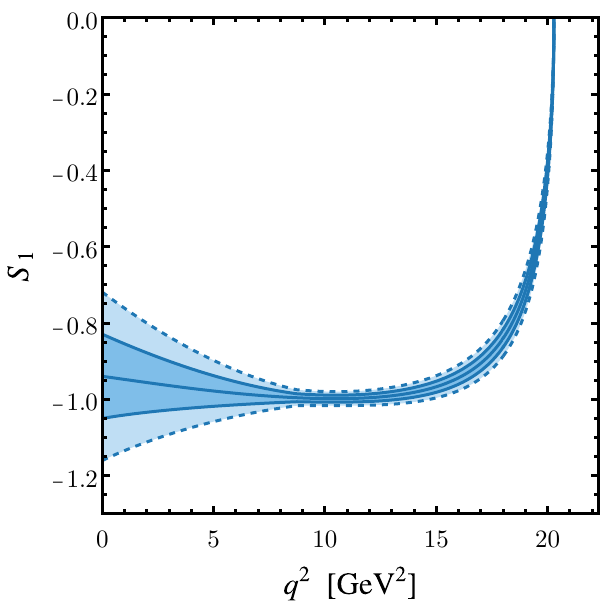} \qquad
\includegraphics[width=0.46\textwidth]{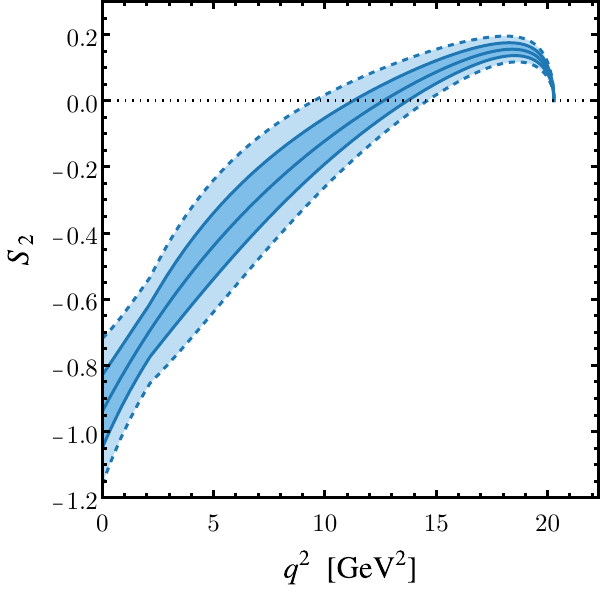} \\[12pt]
\includegraphics[width=0.46\textwidth]{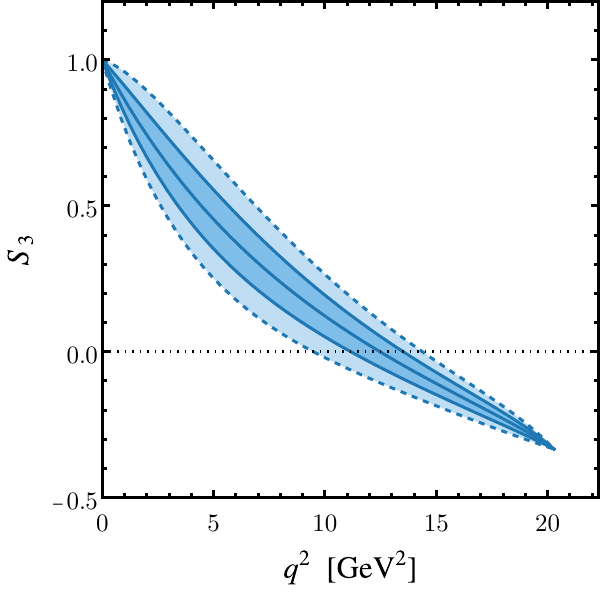} \qquad
\includegraphics[width=0.46\textwidth]{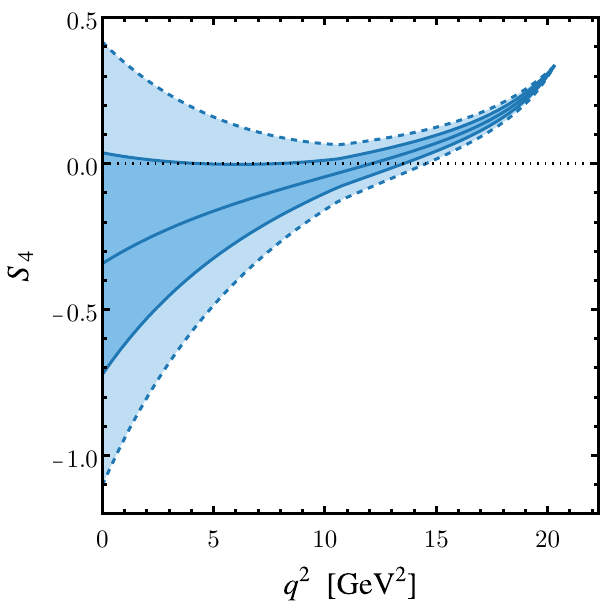}
\caption{The angular observables $S_1$ (top left), $S_2$ (top right), $S_3$ (bottom left), and $S_4$ (bottom right) as a function of $q^2$ in the SM. The dark and light blue bands show the $1\sigma$ and $2\sigma$ uncertainties.}
\label{fig:Si_SM}
\end{figure}

In figure~\ref{fig:Si_SM}, we show the $q^2$ dependence of the angular observables $S_1$, $S_2$, $S_3$, and $S_4$ in the SM. The dark and light blue bands correspond to the $1\sigma$ and $2\sigma$ uncertainties coming from the form factors.
We note that these observables obey endpoint relations~\cite{Hiller:2021zth, Iguro:2026rxw} at $q^2_\text{max} = (m_{\Lambda_b} - m_\Lambda)^2$ that hold in the SM, and also in any new physics scenario described by the effective Hamiltonian in~\eqref{eq:Heff}. One has
\begin{equation}
S_1(q^2_\text{max}) = S_2(q^2_\text{max}) = 0 ~, \quad S_3(q^2_\text{max}) = - \frac{1}{3} ~,\quad S_4(q^2_\text{max}) = +\frac{1}{3}~.
\end{equation}

As already anticipated above, the angular observables $S_2$, $S_3$, and $S_4$ have zero crossings in $q^2$, which are determined by the following conditions
\begin{eqnarray}
q_0^2(S_2): && \quad \frac{\mathcal F_0^{VA}}{\mathcal F_\perp^{VA}} = 1 ~,  \\
q_0^2(S_3): && \quad \frac{\big| C_R + C_L \big|^2 \big( \mathcal F_0^V - \mathcal F_\perp^V \big)}{\big| C_R - C_L \big|^2 \big( \mathcal F_\perp^A - \mathcal F_0^A \big)} = 1 ~,  \\
q_0^2(S_4): && \quad \frac{\big| C_R + C_L \big|^2  \mathcal F_0^V }{\big| C_R - C_L \big|^2  \mathcal F_0^A} = 1 ~.
\end{eqnarray}
The zero crossing of $S_2$ is independent of Wilson coefficients and fully determined by combinations of form factors. As pointed out already in~\cite{Altmannshofer:2025eor} this zero crossing is therefore independent of new physics and can serve as a test of form factor calculations. The zero crossings of $S_3$, and $S_4$ do in general depend on new physics Wilson coefficients. In the SM, we find the values
\begin{eqnarray}
q_0^2(S_2^\text{SM}) &=& (12.6 \pm 1.2)\,\text{GeV}^2 ~, \\
q_0^2(S_3^\text{SM}) &=& (12.5 \pm 1.2)\,\text{GeV}^2 ~, \\
q_0^2(S_4^\text{SM}) &=& (12.1 \pm 3.8)\,\text{GeV}^2~.  
\end{eqnarray}
The uncertainty of the $S_4$ zero crossing is sizable, as the observable $S_4$ is close to zero over a broad range of $q^2$ values, see figure~\ref{fig:Si_SM}.

\subsection{New Physics Benchmark Scenarios} \label{sec:NP}

\begin{figure}[tb]
\centering
\includegraphics[width=0.46\textwidth]{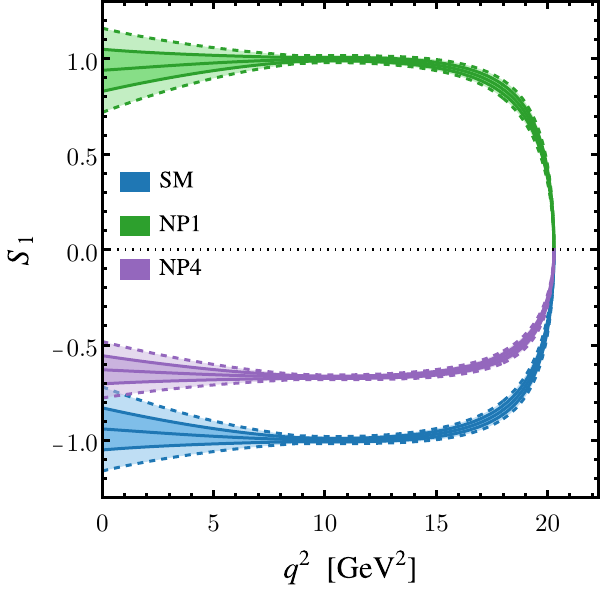} \qquad
\includegraphics[width=0.46\textwidth]{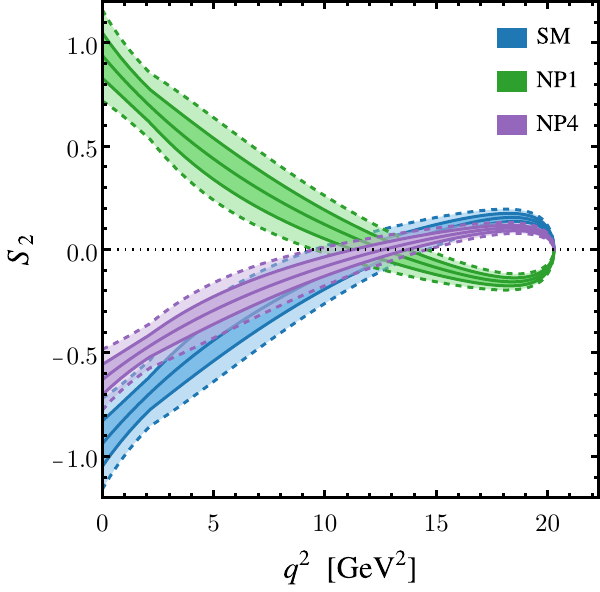} \\[12pt]
\includegraphics[width=0.46\textwidth]{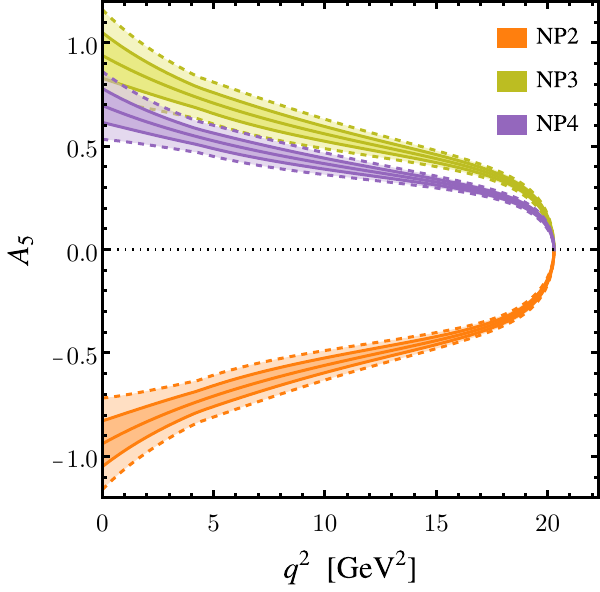}
\caption{The angular observables $S_1$ (top left), $S_2$ (top right), and $A_5$ (bottom) as a function of $q^2$ in the SM (blue) and the four new physics benchmarks defined in equations~\eqref{eq:benchmark1} - \eqref{eq:benchmark4} (green, orange, yellow, purple). The dark and light bands show the $1\sigma$ and $2\sigma$ uncertainties.}
\label{fig:angular_NP}
\end{figure}

To illustrate the new physics sensitivity of the angular observables, we consider several representative benchmark scenarios. They are chosen such that $|C_L+C_R| \simeq |C_L-C_R| \simeq |C_L^\text{SM}|$, so that all standard mesonic $b\to s\nu\bar\nu$ observables remain, to an excellent approximation, identical to their SM predictions. As discussed in the previous section~\ref{sec:Wilson}, this also implies that the $\Lambda_b\to\Lambda\nu\bar\nu$ branching ratio and the angular observables $S_3$ and $S_4$ are essentially SM-like. The benchmark points are
\begin{eqnarray} \label{eq:benchmark1}
 \text{(NP1)}: \qquad && C_L = 0 ~,\quad C_R = C_L^\text{SM} ~, \\  \label{eq:benchmark2}
 \text{(NP2)}: \qquad && C_L = C_L^\text{SM}/\sqrt{2} ~,\quad C_R = i\,C_L^\text{SM}/\sqrt{2} ~, \\  \label{eq:benchmark3}
 \text{(NP3)}: \qquad && C_L = C_L^\text{SM}/\sqrt{2} ~,\quad C_R = -i\,C_L^\text{SM}/\sqrt{2} ~, \\  \label{eq:benchmark4}
 \text{(NP4)}: \qquad && C_L = 0.9\,C_L^\text{SM} ~,\quad C_R = 0.4\,i\,C_L^\text{SM} ~.
\end{eqnarray}

Figure~\ref{fig:angular_NP} shows the corresponding predictions for the angular observables $S_1$, $S_2$, and $A_5$. A general feature is that deviations from the SM require the presence of right-handed currents. If new physics modifies only the left-handed coefficient $C_L$, all angular observables remain unchanged, as they depend only on ratios of Wilson coefficient combinations that are insensitive to an overall rescaling of $C_L$. Consequently, the angular observables provide direct probes of right-handed interactions in $b\to s\nu\bar\nu$ transitions.

The observables $S_1$ and $S_2$ are governed by the combination $|C_R|^2-|C_L|^2$. They are therefore maximally sensitive to right-handed currents and, in scenario (NP1) shown by the green bands, reverse sign relative to their SM predictions. In the benchmark scenarios (NP2) and (NP3), one has $|C_L|=|C_R|$, implying $S_1=S_2=0$ (not shown in the figure). In the more realistic scenario (NP4) shown in purple, the deviations are more moderate, but remain clearly visible, particularly for $S_1$, whose theoretical uncertainty is very small.

The observable $A_5$ plays a unique role. It is a $T$-odd CP asymmetry that is directly proportional to $\mathrm{Im}(C_LC_R^*)$ and therefore provides clean access to the relative CP-violating phase between the short-distance Wilson coefficients. Accordingly, $A_5$ is zero in the SM, while it reaches its maximal magnitude in the benchmark scenarios (NP2) and (NP3), shown by the orange and yellow bands. Even in the more conservative benchmark (NP4) shown in purple, the predicted asymmetry remains sizable, illustrating the excellent sensitivity of polarized $\Lambda_b\to\Lambda\nu\bar\nu$ decays to CP-violating new physics.

\section{Prospects for New Physics Searches} \label{sec:prospects}

New physics contributions to the $b \to s \nu\bar\nu$ transition can modify both the branching ratio and the angular observables of the decay $\Lambda_b \to \Lambda(\to p\pi^-)\nu\bar\nu$, as well as their $q^2$ distributions. The expressions for the integrated branching ratio and angular observables in the presence of new physics are
\begin{eqnarray}
 \text{BR}(\Lambda_b \to \Lambda \nu\bar\nu) &=& \frac{\text{BR}(\Lambda_b \to \Lambda \nu\bar\nu)_\text{SM}}{|C_L^\text{SM}|^2} \Big( r \big| C_L + C_R \big|^2  + (1-r) \big| C_L - C_R \big|^2  \Big) ~, \\
 \langle S_1 \rangle &=& \frac{  \langle S_1^\text{SM} \rangle \Big( \big| C_L \big|^2 - \big| C_R \big|^2 \Big) }{r \big| C_L + C_R \big|^2  + (1-r) \big| C_L - C_R \big|^2} ~, \\
 \langle S_2 \rangle &=& \frac{  \langle S_2^\text{SM} \rangle \Big( \big| C_L \big|^2 - \big| C_R \big|^2 \Big) }{r \big| C_L + C_R \big|^2  + (1-r) \big| C_L - C_R \big|^2} ~, \\
 \langle S_3 \rangle &=& \frac{  \langle S_3^\text{SM} \rangle \Big( s \big| C_L + C_R \big|^2  + (1-s) \big| C_L - C_R \big|^2 \Big) }{r \big| C_L + C_R \big|^2  + (1-r) \big| C_L - C_R \big|^2} ~, \\
 \langle S_4 \rangle &=& \frac{  \langle S_4^\text{SM} \rangle \Big( t \big| C_L + C_R \big|^2  + (1-t) \big| C_L - C_R \big|^2 \Big) }{r \big| C_L + C_R \big|^2  + (1-r) \big| C_L - C_R \big|^2} ~, \\
 \langle A_5 \rangle &=& \frac{ - \big( \langle S_1^\text{SM} \rangle + \langle S_2^\text{SM} \rangle  \big) ~ \text{Im} \big( C_L^* C_R \big)}{r \big| C_L + C_R \big|^2  + (1-r) \big| C_L - C_R \big|^2} ~.
\end{eqnarray}
The coefficients $r$, $s$, and $t$ are given by
\begin{eqnarray}
r &=& \frac{\int_{q^2_\text{min}}^{q^2_\text{max}} dq^2 ~ \lambda^{\frac{1}{2}}~ \big( \mathcal F_0^V + \mathcal F_\perp^V \big)}{\int_{q^2_\text{min}}^{q^2_\text{max}} dq^2 ~ \lambda^{\frac{1}{2}} ~\big( \mathcal F_0^V + \mathcal F_\perp^V + \mathcal F_0^A + \mathcal F_\perp^A \big) } \simeq 0.408 ~, \\
s &=& \frac{\int_{q^2_\text{min}}^{q^2_\text{max}} dq^2 ~ \lambda^{\frac{1}{2}}~ \big( \mathcal F_0^V - \mathcal F_\perp^V \big)}{\int_{q^2_\text{min}}^{q^2_\text{max}} dq^2 ~ \lambda^{\frac{1}{2}} ~\big( \mathcal F_0^V - \mathcal F_\perp^V + \mathcal F_0^A - \mathcal F_\perp^A \big) } \simeq 2.998 ~, \\
t &=& \frac{\int_{q^2_\text{min}}^{q^2_\text{max}} dq^2 ~ \lambda^{\frac{1}{2}}~  \mathcal F_0^V }{\int_{q^2_\text{min}}^{q^2_\text{max}} dq^2 ~ \lambda^{\frac{1}{2}} ~\big( \mathcal F_0^V - \mathcal F_0^A \big) } \simeq -8.422~.
\end{eqnarray}
where $\lambda^{\frac{1}{2}} = \sqrt{\lambda\left( 1,\frac{m_\Lambda^2}{m_{\Lambda_b}^2}, \frac{q^2}{m_{\Lambda_b}^2}\right)}$. The numerical values for $r$, $s$, and $t$ are obtained by integrating over the full kinematic $q^2$ range using the central values of the form factors.

The sensitivity of $\Lambda_b \to \Lambda(\to p\pi^-)\nu\bar\nu$ to new physics depends both on the theoretical uncertainties of the SM predictions and on the achievable experimental precision. The uncertainty of the branching ratio measurement at FCC-ee was estimated to be approximately $10\%$ in~\cite{Amhis:2023mpj}, and we adopt this value throughout our analysis. At present, no dedicated estimates exist for the experimental precision of the angular observables. 

In our analysis, we take into account the correlated theoretical uncertainties of the SM predictions discussed in section~\ref{sec:SM}, and combine them with assumed experimental uncertainties.
On the experimental side, we consider a $10\%$ uncertainty on the branching ratio and study the projected new physics sensitivity for uncorrelated experimental uncertainties of $\pm 0.1$, $\pm 0.2$, $\pm 0.3$, and $\pm 0.5$ on the angular observables $S_i$ and $A_5$. The theory and experimental uncertainties are combined into a covariance matrix, $C = C_\text{th} + C_\text{exp}$. Throughout this analysis we neglect the dependence of the theoretical uncertainties on the Wilson coefficients.

Assuming that the future measurements agree with the SM predictions, the projected sensitivity is quantified through a $\chi^2$ function
\begin{equation} \label{eq:chi2}
\chi^2(C_L,C_R) = \big( \vec O(C_L,C_R)-\vec O_\text{SM} \big)^T ~C^{-1}~ \big( \vec O(C_L,C_R)-\vec O_\text{SM} \big) ~,
\end{equation}
where $\vec O$ denotes the vector of observables $\vec O= \{ \text{BR}, S_1, S_2, S_3, S_4, A_5 \}$.
The resulting $\chi^2$ is minimized with respect to the Wilson coefficients, or profiled over the remaining Wilson coefficient combinations where appropriate.

For comparison, we also estimate the sensitivity of future mesonic measurements. According to~\cite{Amhis:2023mpj}, the branching ratios of $B \to K\nu\bar\nu$ and $B \to K^*\nu\bar\nu$ could be measured with approximately $1\%$ precision at FCC-ee, and we expect that the longitudinal polarization fraction $F_L(B \to K^*\nu\bar\nu)$ might reach a comparable level of precision. In this case, the ultimate sensitivity is likely limited by theoretical uncertainties. Since a detailed analysis of the mesonic observables is beyond the scope of this work, we use a simplified estimate and assume a combined theory and experimental uncertainty of $5\%$ for $\text{BR}(B \to K\nu\bar\nu)$, $\text{BR}(B \to K^*\nu\bar\nu)$, and $F_L(B \to K^*\nu\bar\nu)$, neglecting correlations among these observables. This comparison is intended only as a qualitative illustration of the relative new physics reach of baryonic and mesonic modes.

\subsection{Projected Sensitivity for Real Wilson Coefficients}

\begin{figure}[tb]
\centering
\includegraphics[width=0.46\textwidth]{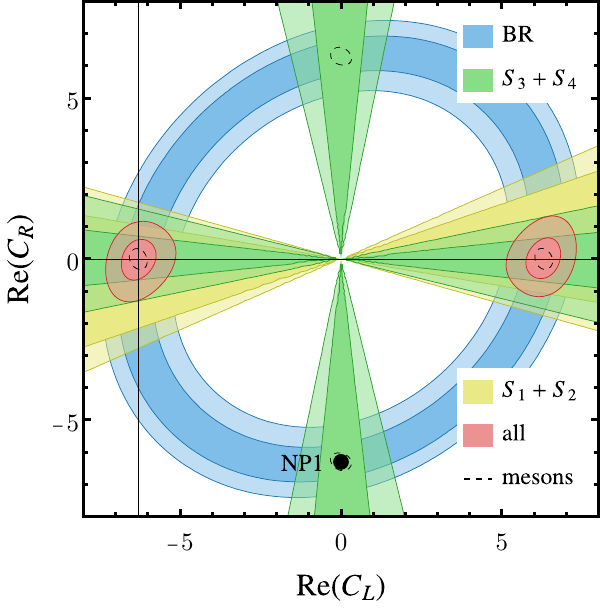} \qquad
\includegraphics[width=0.46\textwidth]{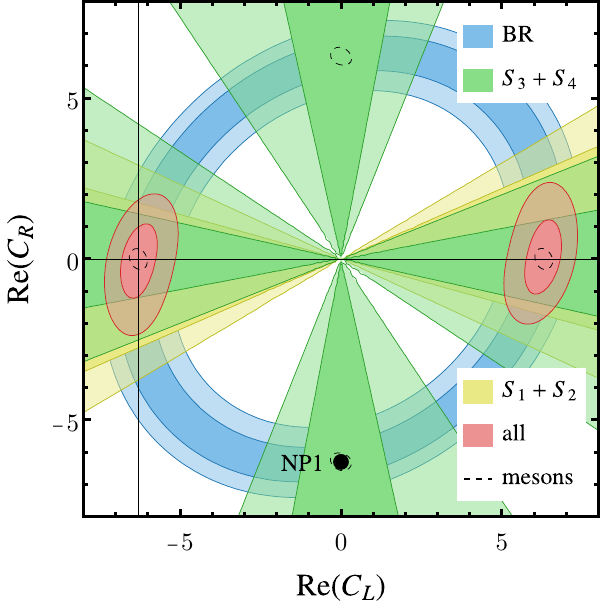} \\[12pt]
\includegraphics[width=0.46\textwidth]{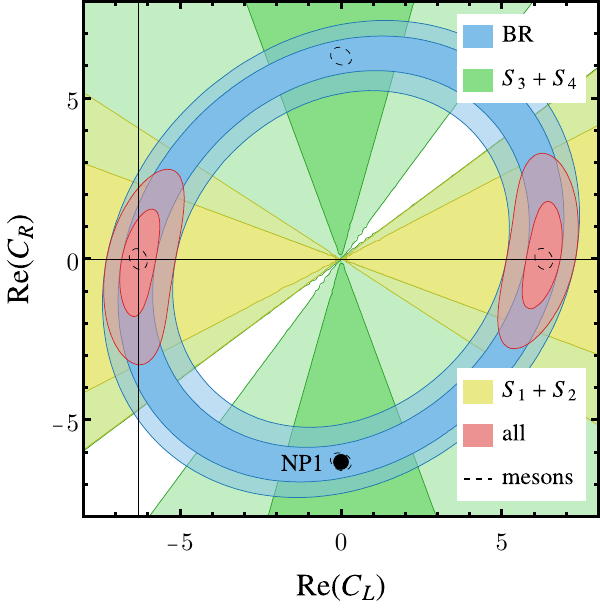} \qquad
\includegraphics[width=0.46\textwidth]{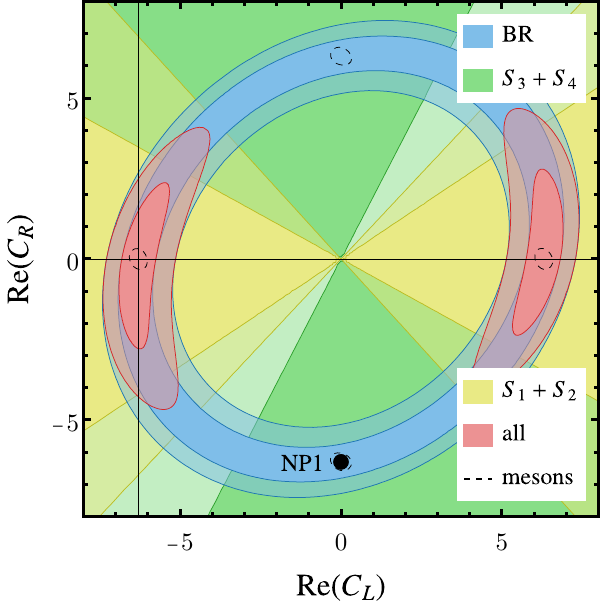}
\caption{Projected sensitivities of $\Lambda_b \to \Lambda (\to p \pi^-) \nu \bar\nu$ observables in the plane of Wilson coefficients $\text{Re}(C_L)$ and $\text{Re}(C_R)$. The imaginary parts of the Wilson coefficients are set to zero. The dark and light shaded regions correspond to $1\sigma$ and $2\sigma$. The black dashed contours show an estimate of the future expected $2\sigma$ sensitivity of meson observables (see text for details). From top left to bottom right the assumed experimental uncertainties on the angular observables are $\pm 0.1$, $\pm 0.2$, $\pm 0.3$, and $\pm 0.5$. The SM point corresponds to $(C_L, C_R) \simeq (-6.3, 0)$. The new physics scenario (NP1) from equation~\eqref{eq:benchmark1} is also shown.}
\label{fig:fit1}
\end{figure}

In figure~\ref{fig:fit1} we show the projected sensitivity of the $\Lambda_b \to \Lambda(\to p\pi^-)\nu\bar\nu$ observables in the plane of the real Wilson coefficients $\mathrm{Re}(C_L)$ and $\mathrm{Re}(C_R)$, with the imaginary parts fixed to zero. Even in this CP-conserving scenario, the baryonic angular observables provide significant additional information. The dark and light shaded regions correspond to the projected $1\sigma$ and $2\sigma$ sensitivities, defined by $\Delta\chi^2=1,4$ for individual observables and $\Delta\chi^2=2.3,6$ for their combination. The black dashed contours show the estimated future $2\sigma$ sensitivity of the mesonic observables as discussed above. From top left to bottom right the panels assume experimental uncertainties of $\pm 0.1$, $\pm 0.2$, $\pm 0.3$, and $\pm 0.5$ for the baryonic angular observables.
The four plots illustrate the gradual evolution of the new physics reach as the experimental precision of the baryonic angular observables is varied.

The observables $S_1$ and $S_2$, shown in yellow, depend on the same combination of Wilson coefficients and are therefore combined. Of the two, $S_1$ provides the stronger constraint. Likewise, $S_3$ and $S_4$, shown in green, depend on similar combinations of Wilson coefficients and are also combined. Since $A_5$ vanishes identically for real Wilson coefficients, it does not contribute in this scenario.

The combination of the branching ratio (shown in blue) with $S_3$ and $S_4$ selects four disconnected regions of parameter space, consistent with the information provided by the mesonic observables, although with less precision. Neither the mesonic observables nor the combination of $\Lambda_b \to \Lambda\nu\bar\nu$ branching ratio, $S_3$, and $S_4$ can distinguish between these four solutions, corresponding approximately to $C_L \simeq \pm C_L^\text{SM}$ with $C_R \simeq 0$ or $C_L \simeq 0$ with $C_R \simeq \pm C_L^\text{SM}$. One of the latter two corresponds to the benchmark scenario (NP1) introduced in section~\ref{sec:NP}. The observables $S_1$ and $S_2$, however, lift this degeneracy. Once all observables are combined, shown in red, only the solutions with $C_L \simeq \pm C_L^\text{SM}$, $C_R \simeq 0$ remain. This conclusion is very robust: it is clearly visible for $\pm 0.1$ experimental uncertainties and continues to hold even if the uncertainties of the angular observables increase to $\pm 0.5$.

\subsection{Projected Sensitivity for Complex Wilson Coefficients}

\begin{figure}[tb]
\centering
\includegraphics[width=0.46\textwidth]{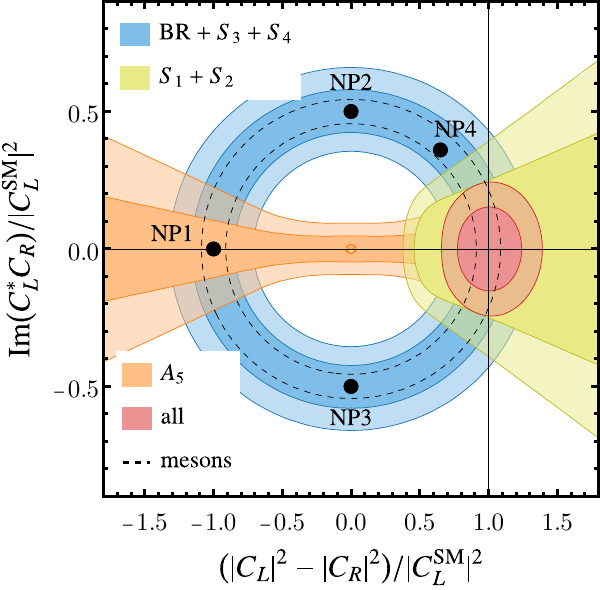} \qquad
\includegraphics[width=0.46\textwidth]{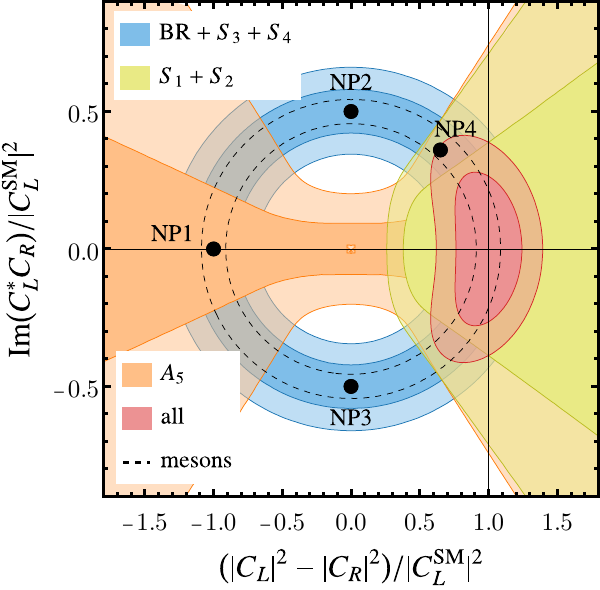} \\[12pt]
\includegraphics[width=0.46\textwidth]{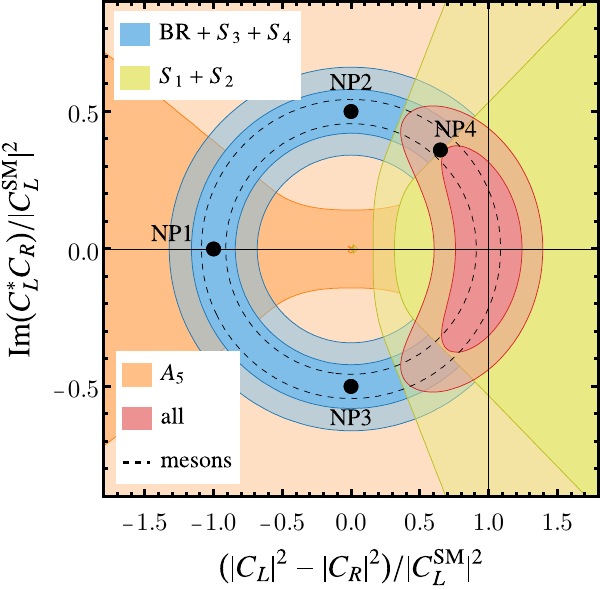} \qquad
\includegraphics[width=0.46\textwidth]{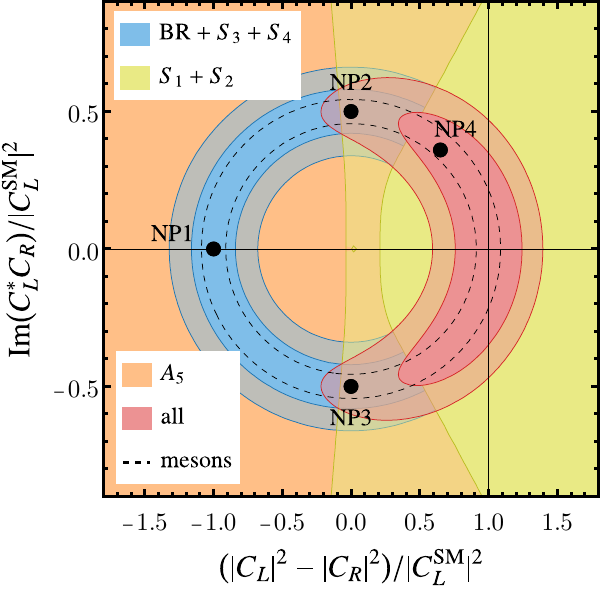}
\caption{Projected sensitivities of $\Lambda_b \to \Lambda (\to p \pi^-) \nu \bar\nu$ observables in the plane of Wilson coefficient combinations $(|C_L|^2 - |C_R|^2)/|C_L^\text{SM}|^2$ and $\text{Im}(C_L^* C_R)/|C_L^\text{SM}|^2$. The third independent combination of Wilson coefficients is profiled to achieve agreement with meson observables as well as possible. The dark and light shaded regions correspond to $1\sigma$ and $2\sigma$. The black dashed contours show an estimate of the future expected $2\sigma$ sensitivity of meson observables (see text for details). From top left to bottom right the assumed experimental uncertainties on the angular observables are $\pm 0.1$, $\pm 0.2$, $\pm 0.3$, and $\pm 0.5$. The SM point corresponds to $(1, 0)$. The new physics scenarios (NP1)--(NP4) from equations~\eqref{eq:benchmark1}--\eqref{eq:benchmark4} are also shown.}
\label{fig:fit3}
\end{figure}

In figure~\ref{fig:fit3} we extend the analysis to complex Wilson coefficients and present the projected sensitivity in the plane spanned by $|C_L|^2-|C_R|^2$ and $\text{Im}(C_L^*C_R)$, normalized to the absolute value squared of the SM Wilson coefficient $|C_L^\text{SM}|^2$, such that the SM prediction is located at $(1,0)$. The third independent combination of Wilson coefficients is profiled at each point in the plots in order to maximize the agreement with the mesonic observables. As before, the dark and light shaded regions correspond to the projected $1\sigma$ and $2\sigma$ sensitivities, while the black dashed contours illustrate the expected $2\sigma$ reach of the mesonic observables. From top left to bottom right the panels assume experimental uncertainties of $\pm 0.1$, $\pm 0.2$, $\pm 0.3$, and $\pm 0.5$ for the baryonic angular observables. As anticipated in section~\ref{sec:Wilson}, the mesonic observables constrain the Wilson coefficients to a ring-shaped region. 

The combination of the $\Lambda_b \to \Lambda\nu\bar\nu$ branching ratio with the observables $S_3$ and $S_4$, shown in blue, follows the same ring-shaped structure preferred by the mesonic observables, albeit with a larger uncertainty. The observables $S_1+S_2$, shown in yellow, and $A_5$, shown in orange, provide complementary information and together isolate a region around the SM point as the only viable solution. Also indicated in the plot are the benchmark scenarios (NP1)--(NP4) introduced in equations~\eqref{eq:benchmark1}--\eqref{eq:benchmark4}. Assuming a $\pm 0.1$ experimental precision for the angular observables, all benchmark scenarios can be distinguished from the SM. If the precision deteriorates to $\pm 0.5$, however, the sensitivity of $A_5$ is largely lost. In this case only benchmark case (NP1) remains distinguishable, primarily through the observable $S_1$, while the other benchmark scenarios become compatible with the projected measurements at the $2\sigma$ level.

\begin{figure}[tb]
\centering
\includegraphics[width=0.46\textwidth]{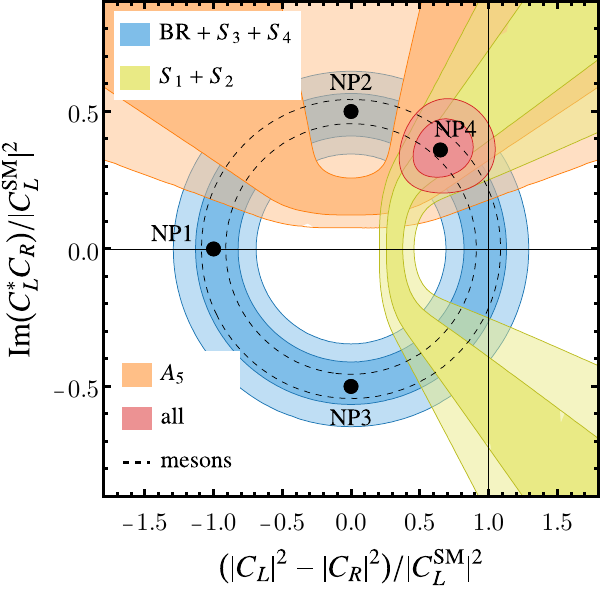} \qquad
\includegraphics[width=0.46\textwidth]{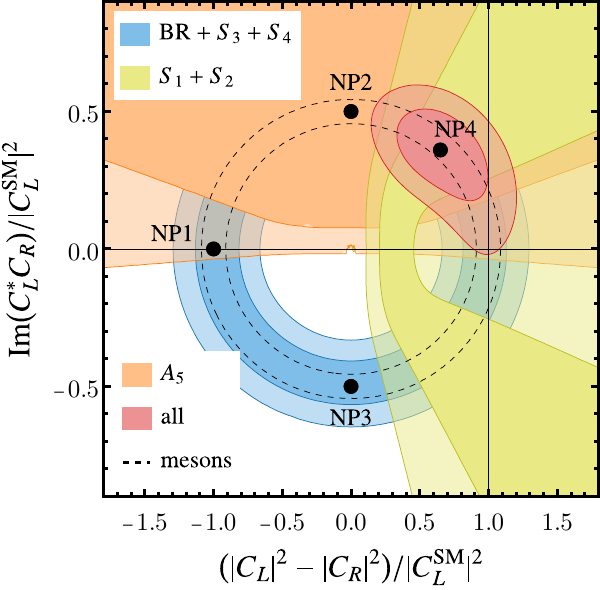} \\[12pt]
\includegraphics[width=0.46\textwidth]{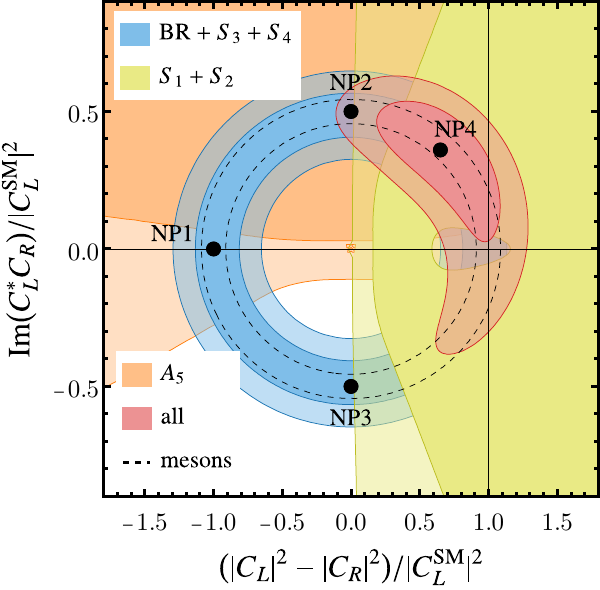} \qquad
\includegraphics[width=0.46\textwidth]{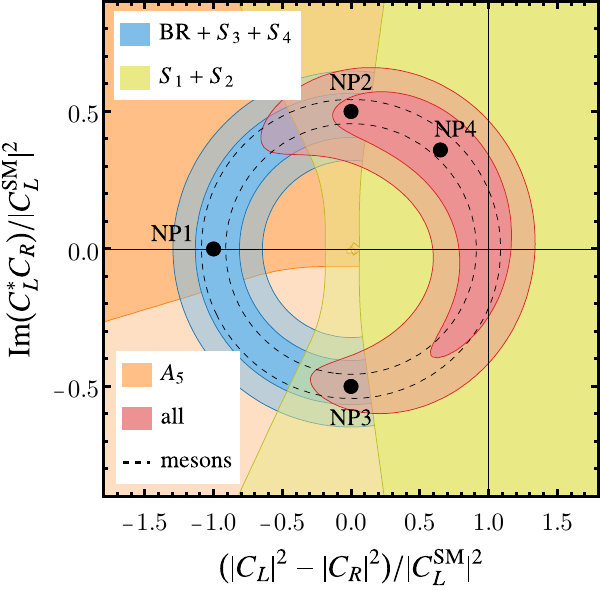}
\caption{As figure~\ref{fig:fit3}, but with assumed experimental central values that correspond to the new physics benchmark point (NP4) from equation~\eqref{eq:benchmark4}. From top left to bottom right the assumed experimental uncertainties on the angular observables are $\pm 0.1$, $\pm 0.2$, $\pm 0.3$, and $\pm 0.5$.}
\label{fig:NP4_sensitivity}
\end{figure}

To complement our baseline projections, which assume the SM predictions as the central values of future measurements, we also explore a scenario in which the experimental results align with the new physics benchmark point (NP4) from equation~\eqref{eq:benchmark4}. The resulting Wilson coefficient constraints for various levels of experimental precision on the angular observables are collected in figure~\ref{fig:NP4_sensitivity}. When assuming a precise experimental measurement with an uncertainty of $\pm 0.1$ on the observables, the combined fit region can be clearly distinguished from the SM expectation. If the experimental uncertainty increases to $\pm 0.2$, the best-fit region expands such that its $2\sigma$ boundary just touches the SM point. For larger uncertainties ($\gtrsim \pm 0.3$), the best-fit region becomes fully compatible with the SM within $2\sigma$. This demonstrates that while coarse measurements remain valuable, achieving an experimental uncertainty of $\mathcal{O}(0.1-0.2)$ is important for cleanly resolving NP scenarios like (NP4) from the SM baseline.

\bigskip
In~\cite{Altmannshofer:2025eor} it was shown that the new physics sensitivity of the spin forward-backward asymmetry $A_\text{FB}^\uparrow$ (which corresponds to the observable $S_2$ in our notation, see equation~\eqref{eq:AFB}) can be significantly enhanced by measuring it in two $q^2$ bins, one below and one above its zero crossing. We expect a similar improvement for the other two observables with zero crossings, $S_3$ and $S_4$. We have therefore investigated the sensitivity gain from splitting the $q^2$ region into two bins. Since the zero crossings of $S_2$, $S_3$, and $S_4$ occur at very similar values of $q^2$, we use a common binning for all observables, with the bin boundary given by the zero crossing of $S_2$, $q^2_0(S_2) \simeq 12.6\,\text{GeV}^2$.

The results of this exercise are collected in appendix~\ref{app:q2_binning}. Figure~\ref{fig:fit1_q2} is analogous to figure~\ref{fig:fit1} and shows projected sensitivities in the plane of Wilson coefficients $\text{Re}(C_L)$ and $\text{Re}(C_R)$, with the imaginary parts of the Wilson coefficients set to zero. Compared to a single wide $q^2$ bin, we observe a visible improvement in the sensitivity, although it is less pronounced than in~\cite{Altmannshofer:2025eor}, where only a single angular observable was considered. The combination of all angular observables already provides good sensitivity without $q^2$ binning. Splitting the $q^2$ region further improves the sensitivity, but does not lead to a qualitative change in the resulting constraints.

The same behavior is observed in figure~\ref{fig:fit3_q2}, which is analogous to figure~\ref{fig:fit3} and shows projected sensitivities in the plane of the Wilson coefficient combinations $(|C_L|^2 - |C_R|^2)/|C_L^\text{SM}|^2$ and $\text{Im}(C_L^* C_R)/|C_L^\text{SM}|^2$. The third independent combination of Wilson coefficients is profiled over, subject to the constraints from meson observables. The allowed regions in the new physics parameter space shrink visibly when the $q^2$ binning is included but the qualitative picture remains unchanged.

\section{Summary and Conclusions} \label{sec:conclusions}

In this work, we have performed a detailed analysis of the rare baryonic decay $\Lambda_b \rightarrow \Lambda(\rightarrow p\pi^-)\nu\bar{\nu}$ in the SM and beyond. While mesonic rare di-neutrino decays ($B \rightarrow K\nu\bar{\nu}$, $B \rightarrow K^*\nu\bar{\nu}$, $B_s \rightarrow \phi\nu\bar{\nu}$) serve as the primary probes of $b \rightarrow s\nu\bar{\nu}$ transitions at Belle II and FCC-ee, they are inherently limited. Even precise measurements of their branching ratios and angular distributions leave a continuous degeneracy in the underlying short-distance parameters, as they are sensitive only to the two combinations of Wilson coefficients $|C_L + C_R|$ and $|C_L - C_R|$.

We have demonstrated that the rare decays of polarized $\Lambda_b$ baryons, available in large quantities at the $Z$ pole, provide crucial complementary information that overcomes these limitations. Our main conclusions are summarized as follows:

\begin{itemize}
    \item The differential decay distribution of $\Lambda_b \rightarrow \Lambda(\rightarrow p\pi^-)\nu\bar{\nu}$ is completely described by three hadronic angles $(\theta, \theta_p, \phi_p)$ and yields five independent angular observables: $S_1, S_2, S_3, S_4$, and $A_5$. The observable $S_1$ corresponds to the hadronic forward-backward asymmetry $A_{\text{FB}}^\Lambda$ and is accessible even for unpolarized $\Lambda_b$ samples. In contrast, $S_2, S_3, S_4$, and $A_5$ explicitly require the initial $\Lambda_b$ longitudinal polarization $\mathcal{P}_{\Lambda_b}$ inherited from $Z \to b\bar{b}$ production.
    
    \item The observables $S_1$ and $S_2$ are governed by the combination $|C_R|^2 - |C_L|^2$, making them uniquely sensitive to right-handed currents and capable of breaking discrete parameter degeneracies present in mesonic channels.
    
    \item The observable $A_5$ is a $T$-odd CP asymmetry directly proportional to $\text{Im}(C_L C_R^*)$, offering a direct probe of short-distance CP violation in $b \rightarrow s\nu\bar{\nu}$ transitions without the need for flavor tagging or time-dependent measurements.
    
    \item The observables $S_2, S_3$, and $S_4$ feature zero crossings in $q^2$. Notably, the zero crossing of $S_2$ at $q_0^2(S_2) = (12.6 \pm 1.2)\,\text{GeV}^2$ is completely independent of Wilson coefficients, providing in principle a clean test of hadronic form factor determinations.
    
    \item We have monitored the sensitivity to new physics as a function of the experimental uncertainty on the angular observables. We find that even a modest experimental measurement of the angular observables with an uncertainty as large as $\pm 0.5$ is still highly valuable and capable of breaking parameter degeneracies and ruling out representative benchmark scenarios of new physics.
    
    \item Projected sensitivity fits at FCC-ee demonstrate that while mesonic observables constrain the short-distance couplings to a ring in the $(|C_L|^2 - |C_R|^2, \text{Im}(C_L^* C_R))$ plane, adding the baryonic angular observables $S_1, S_2$, and $A_5$ allow a complete determination of the three physical parameters $C_L$, $C_R$, and $\text{Arg}(C_R/C_L)$, up to the unobservable common phase.
\end{itemize}

Measuring $\Lambda_b \rightarrow \Lambda(\rightarrow p\pi^-)\nu\bar{\nu}$ at future $Z$-pole facilities like FCC-ee is essential for a comprehensive $b \rightarrow s \nu \bar{\nu}$ program, providing an important tool to determine the chiral nature and CP-violating properties of potential new physics in rare decays.

\section*{Acknowledgments}

The research of WA is supported by the U.S. Department of Energy grant number DE-SC0010107. WA is grateful to the Mainz Institute for Theoretical Physics (MITP) of the Cluster of Excellence PRISMA++ (Project ID 390831469) for its hospitality and its support during parts of this work. SR acknowledges support from the Department of Atomic Physics, Institute of Physics and Astronomy, Faculty of Science, ELTE, and from the research project PZ1104/25 (grant file no. 151398), entitled ``Semi-leptonic decays of baryons: a promising probe for new physics''.

\appendix
\section{Definition of Angular Variables} \label{app:angles}

In this appendix we give a precise definition of the angular variables used throughout this work. We consider the cascade decay
\begin{equation}
\Lambda_b \to \Lambda(\to p\pi^-) \,V^*(\to\nu\bar\nu) ~,
\end{equation}
where $V^*$ denotes the effective current creating the $\nu\bar\nu$ pair.

We begin in the $\Lambda_b$ rest frame. The momentum of the outgoing $\Lambda$ baryon, $\vec p_\Lambda$, together with the polarization vector of the $\Lambda_b$ baryon, $\hat s_{\Lambda_b}$, define an orthonormal coordinate system
\begin{equation}
\hat z_\Lambda = \frac{\vec p_\Lambda}{|\vec p_\Lambda|} ~,\quad \hat y_\Lambda = \frac{\hat s_{\Lambda_b}\times\hat z_\Lambda}{|\hat s_{\Lambda_b}\times\hat z_\Lambda|} ~,\quad \hat x_\Lambda = \hat y_\Lambda\times\hat z_\Lambda ~.
\end{equation}

The helicity frame of the $\Lambda$ is obtained by boosting into the $\Lambda$ rest frame along the direction of $\vec p_\Lambda$. Since the boost is parallel to the $\hat z_\Lambda$ axis, the three basis vectors remain unchanged.
The proton helicity angles are then defined by
\begin{equation}
\cos\theta_p = \frac{\vec p_p\cdot\hat z_\Lambda} {|\vec p_p|} ~,\quad \cos\phi_p = \frac{\vec p_p\cdot\hat x_\Lambda}{|\vec p_p|\sin\theta_p} ~,\quad \sin\phi_p = \frac{\vec p_p\cdot\hat y_\Lambda}{|\vec p_p|\sin\theta_p} ~,
\end{equation}
where $\vec p_p$ denotes the proton momentum in the $\Lambda$ rest frame.

With these conventions, the decay $\Lambda \to p \pi^-$ is described by the Wigner $D$ function
\begin{equation}
D^{1/2\,*}_{\lambda_\Lambda,\,\lambda_p} (\phi_p,\theta_p,-\phi_p)~.
\end{equation}

The helicity frame of the virtual current $V^*$ is defined analogously. In the $\Lambda_b$ rest frame we introduce
\begin{equation}
\hat z_V = \frac{\vec p_V}{|\vec p_V|} ~,\quad \hat y_V = \frac{\hat s_{\Lambda_b}\times\hat z_V}{|\hat s_{\Lambda_b}\times\hat z_V|} ~,\quad \hat x_V = \hat y_V\times\hat z_V ~,
\end{equation}

Since the $\Lambda$ and $V^*$ are back-to-back in the $\Lambda_b$ rest frame, $\vec p_\Lambda = -\vec p_V$, the two helicity bases satisfy
\begin{equation}
\hat z_V=-\hat z_\Lambda ~,\quad
\hat y_V=-\hat y_\Lambda ~,\quad
\hat x_V=\hat x_\Lambda~.
\end{equation}

After boosting into the rest frame of the neutrino pair, the helicity angles of the antineutrino are defined by
\begin{equation}
\cos\theta_{\bar\nu} = \frac{\vec p_{\bar\nu}\cdot\hat z_V}{|\vec p_{\bar\nu}|} ~,\quad \cos\phi_{\bar\nu} = \frac{\vec p_{\bar\nu}\cdot\hat x_V}{|\vec p_{\bar\nu}|\sin\theta_{\bar\nu}} ~,\quad \sin\phi_{\bar\nu} = \frac{\vec p_{\bar\nu}\cdot\hat y_V}{|\vec p_{\bar\nu}|\sin\theta_{\bar\nu}} ~,
\end{equation}
where $\vec p_{\bar\nu}$ is evaluated in the $V^*$ rest frame. 

With these conventions, the decay $V^*\to\nu\bar\nu$ is described by the Wigner $D$ function
\begin{equation}
D^{1\,*}_{\lambda_V,\,\lambda_{\bar\nu}-\lambda_\nu} (\phi_{\bar\nu},\theta_{\bar\nu},-\phi_{\bar\nu}) ~.
\end{equation}

\section{Wigner D Functions} \label{app:wigD}
The action of an active rotation $R(\alpha,\beta,\gamma)$ on angular momentum eigenstates, defined with respect to a fixed reference frame, is represented by the operator 
\begin{align}
U(R)=e^{-i\alpha J_z}e^{-i\beta J_y}e^{-i\gamma J_z}~.
\end{align}
The corresponding matrix elements define the Wigner $D$-functions
\begin{eqnarray}
D^j_{m'm}(\alpha,\beta,\gamma)&=&\langle jm'|U(R)|jm\rangle=e^{-im'\alpha}\,d^j_{m'm}(\beta)\,e^{-im\gamma} ~, \\
d^j_{m'm}(\beta) &=& \langle jm'|e^{-i\beta J_y}|jm\rangle~,
\end{eqnarray}
where $d^j_{m'm}(\beta)$ are the Wigner small-$d$ matrix elements.
This convention is followed by Jacob-Wick~\cite{Jacob:1959at} as well as the PDG tables~\cite{ParticleDataGroup:2024cfk}. Mathematica's \texttt{WignerD} is natively defined for passive rotations. In the
Condon--Shortley convention~\cite{Condon:1951} with active rotations used in this work, it should, for example, be rewritten as 
\begin{align}
D^j_{m'm}(\alpha,\beta,\gamma)=e^{-im'\alpha}\,\texttt{WignerD[\{j,m',m\},0,-$\beta$,0]}\,e^{-im\gamma} ~.
\end{align}

\section{Projected Sensitivity with \texorpdfstring{$q^2$}{q2} Binning} \label{app:q2_binning}

\begin{figure}[tb]
\centering
\includegraphics[width=0.46\textwidth]{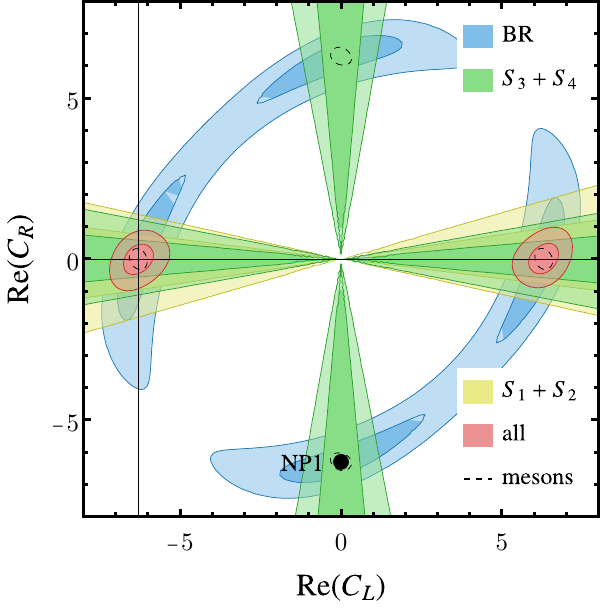} \qquad
\includegraphics[width=0.46\textwidth]{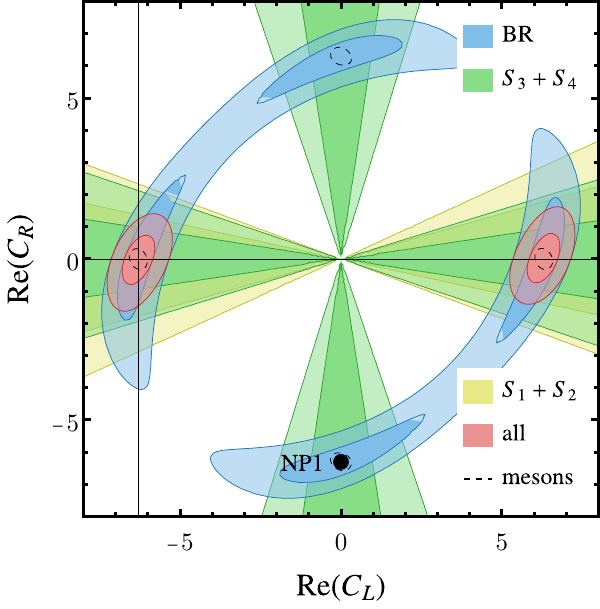} \\[12pt]
\includegraphics[width=0.46\textwidth]{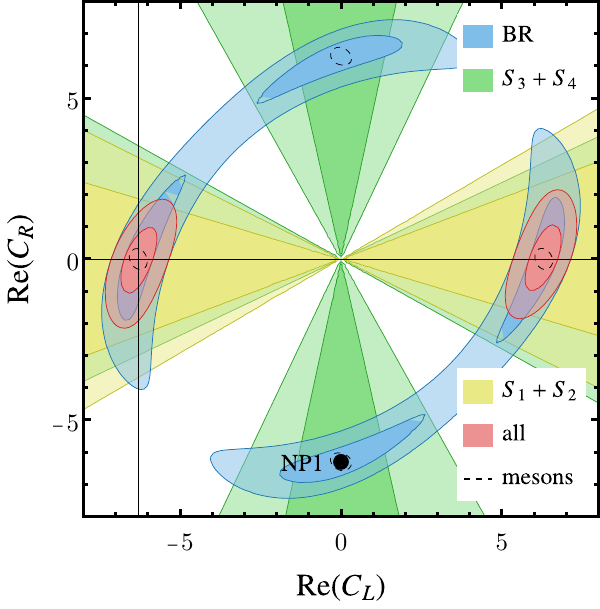} \qquad
\includegraphics[width=0.46\textwidth]{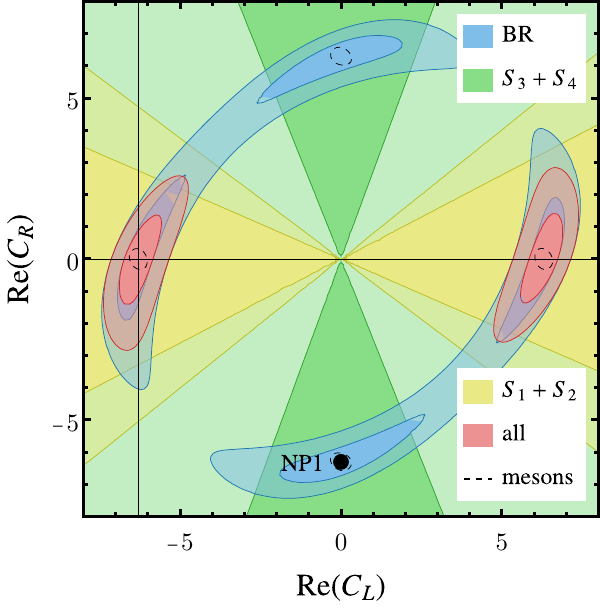}
\caption{Projected sensitivities of $\Lambda_b \to \Lambda (\to p \pi^-) \nu \bar\nu$ observables in the plane of Wilson coefficients $\text{Re}(C_L)$ and $\text{Re}(C_R)$. As figure~\ref{fig:fit1}, but all observables are assumed to be measured in two $q^2$ bins.}
\label{fig:fit1_q2}
\end{figure}

\begin{figure}[tb]
\centering
\includegraphics[width=0.46\textwidth]{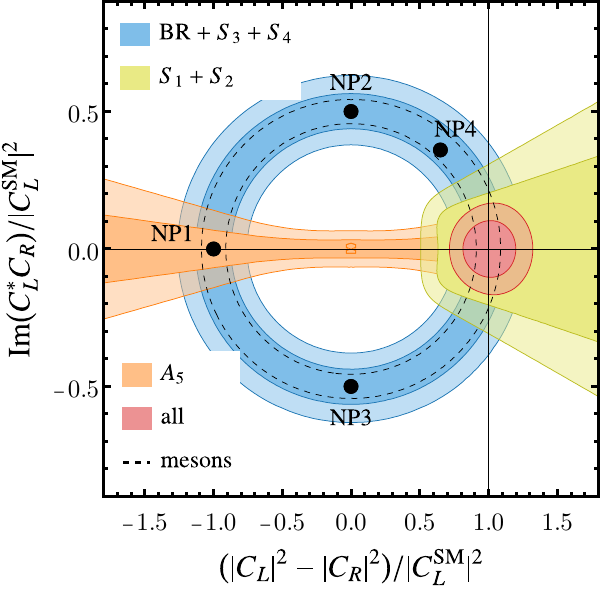} \qquad
\includegraphics[width=0.46\textwidth]{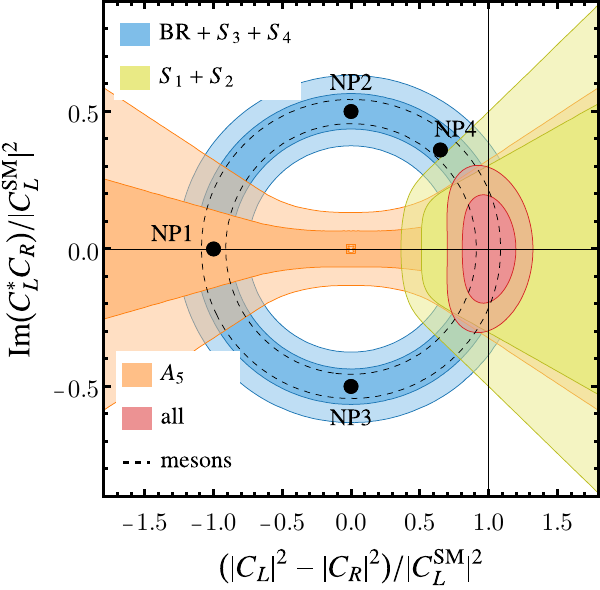} \\[12pt]
\includegraphics[width=0.46\textwidth]{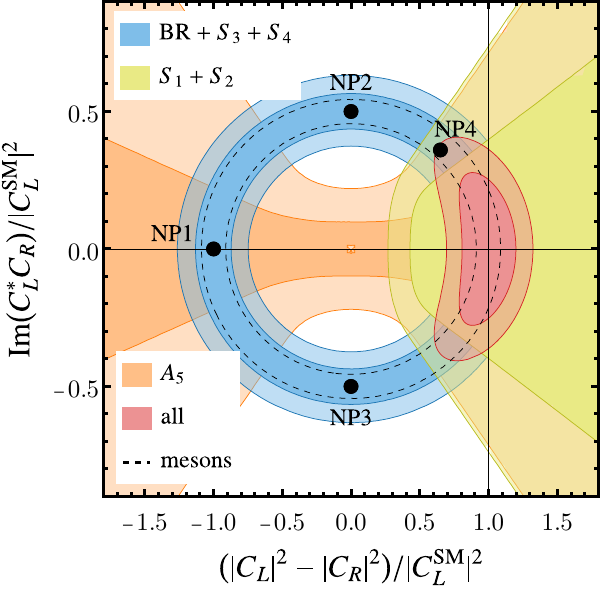} \qquad
\includegraphics[width=0.46\textwidth]{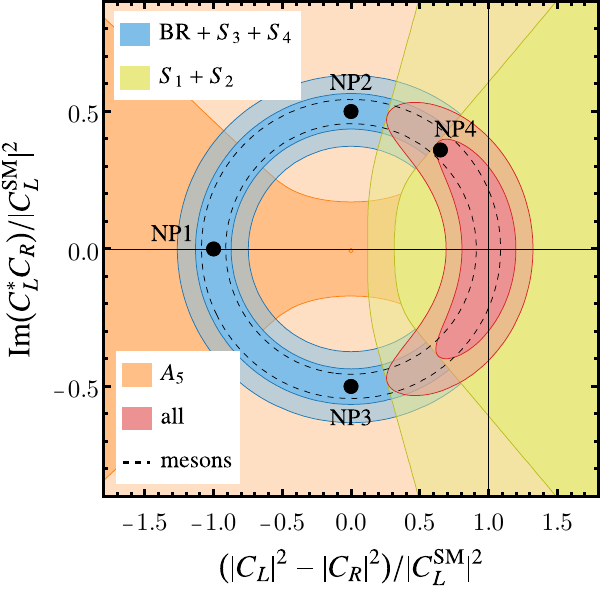}
\caption{Projected sensitivities of $\Lambda_b \to \Lambda (\to p \pi^-) \nu \bar\nu$ observables in the plane of Wilson coefficient combinations $(|C_L|^2 - |C_R|^2)/|C_L^\text{SM}|^2$ and $\text{Im}(C_L^* C_R)/|C_L^\text{SM}|^2$. As figure~\ref{fig:fit3}, but all observables are assumed to be measured in two $q^2$ bins.}
\label{fig:fit3_q2}
\end{figure}

In this appendix we collect the sensitivity projections obtained by splitting the $q^2$ region into a low-$q^2$ bin, $q^2 < 12.6\,\text{GeV}^2$, and a high-$q^2$ bin, $q^2 > 12.6\,\text{GeV}^2$. The same binning is used for all observables, with the boundary chosen according to the zero crossing of $S_2$. We show the resulting constraints for the two Wilson coefficient scenarios considered in the main text, allowing for a direct comparison with the corresponding projections obtained without $q^2$ binning.
Figure~\ref{fig:fit1_q2} is analogous to figure~\ref{fig:fit1}, and figure~\ref{fig:fit3_q2} is analogous to figure~\ref{fig:fit3}.


\bibliographystyle{JHEP}
\bibliography{LBLnunu}

\end{document}